\documentclass[aaspp4]{aastex}
\usepackage{emulateapj5,apjfonts}

\newcommand{\NH}{{$N_{\rm H}$}}

\newcommand{\eps}{ergs s$^{-1}$}
\newcommand{\pcm}{cm$^{-2}$}

\newcommand{\hst}{{\it HST}}
\newcommand{\xmm}{{\it XMM-Newton}}
\newcommand{\chandra}{{\it Chandra}}
\newcommand{\asca}{{\it ASCA}}
\newcommand{\rosat}{{\it ROSAT}}
\newcommand{\Einstein}{{\it Einstein}}

\newcommand{\sax}{{\it BeppoSAX}}

\newcommand{\gtsima}{$\; \buildrel > \over \sim \;$}
\newcommand{\simgt}{\lower.5ex\hbox{\gtsima}}
\newcommand{\ltsima}{$\; \buildrel < \over \sim \;$}
\newcommand{\simlt}{\lower.5ex\hbox{\ltsima}}

\slugcomment{Submitted to {\it The Astrophysical Journal.}}

\lefthead{Terashima et al.}
\righthead{Luminous X-ray Sources in M51}

\begin{document}

\title{The Luminous X-ray Source Population in M51 Observed with {\em Chandra}}

\author{Yuichi Terashima\altaffilmark{1, 2} and
Andrew S. Wilson\altaffilmark{1, 3}
}

\altaffiltext{1}{Astronomy Department, University of Maryland, College Park, 
MD 20742}

\altaffiltext{2}{Institute of Space and Astronautical Science, 3-1-1 Yoshinodai, Sagamihara, Kanagawa 229-8510, Japan}

\altaffiltext{3}{Adjunct Astronomer, Space Telescope Science Institute, 3700
San Martin Drive, Baltimore, MD 21218}

\begin{abstract}

We present the results of two {\it Chandra} observations (separated by
1 year) of the population of X-ray sources in the spiral galaxy M51
(NGC 5194 and NGC 5195). 113 X-ray sources have been detected in an
8\farcm4 $\times$ 8\farcm4 (20.4 $\times$ 20.4 kpc) region, and 84 and
12 of them project within the disks of NGC 5194 and NGC 5195,
respectively. Nine and 28 sources have luminosities exceeding
$1\times10^{39}$ ergs s$^{-1}$ (ultraluminous X-ray sources or ULXs)
and $1\times10^{38}$ ergs s$^{-1}$ in the 0.5 -- 8 keV band,
respectively, assuming they are associated with M51. The number of
ULXs is much higher than found in most normal spiral and elliptical
galaxies.  Most of the X-ray sources and all seven of the ULXs in NGC
5194 are located in, or close to, a spiral arm, suggesting a
connection with recent star formation. The cumulative luminosity
function of the X-ray sources in NGC 5194 with $L$(0.5 -- 8 keV) $>$
10$^{38}$ ergs s$^{-1}$ is well described by a power law N($>L$(0.5 --
8 keV)) $\propto$ $L$(0.5 -- 8 keV)$^{-\alpha}$ with $\alpha$ =
0.91. The X-ray spectra of most of the detected sources are consistent
with a power law with a photon index between 1 and 2, with a few
sources showing harder or softer spectra. The spectra of most ULXs are
consistent with both a power law and a multicolor disk blackbody (MCD)
model, while a power law model is preferable to a MCD model in two
ULXs.  One ULX (NGC 5194 \#69) shows drastic spectral steepening
accompanied by a decline in luminosity by a factor of 3460 in the 2 --
10 keV band between the two observations. This source also exhibited a
possible period of 2.1 hrs in the year 2000 observation. Another ULX
(NGC 5194 \#26) shows strong emission lines from highly ionized
species. The masses of the compact objects and mass accretion rates in
ULXs and other X-ray sources are not well constrained by these
observations. If we adopt a MCD interpretation, their MCD parameters
imply that most of the X-ray sources are stellar mass ($\sim$ 5 -- 10
$M_{\odot}$) black holes accreting near or above the Eddington rate,
although other possibilities (intermediate-mass black holes and
relativistically beamed emission) cannot be excluded.  The power law
sources may, instead, represent Comptonized disk, or nonthermal,
emission.  Two ULXs have very soft spectra; MCD models require
$kT\approx 0.1$ keV. We discuss the possibility that this soft
emission originates in an accretion disk around an intermediate-mass
black hole. We also present a study of the nucleus of, and discrete
sources (including two ULXs) in, the companion galaxy NGC 5195.

\end{abstract}

\keywords{accretion, accretion disks --- galaxies: active 
--- X-rays: binaries --- X-rays: galaxies --- galaxies: Individual (M51)}

\section{Introduction}

The X-ray emission from a galaxy consists of various components
including discrete sources, such as X-ray binaries and supernova
remnants, diffuse hot gas, and an active galactic nucleus (AGN), if
present. One of the most intriguing and puzzling classes of X-ray
source
comprises the so called ultraluminous X-ray
sources (ULXs), whose luminosities ($>10^{39}$ {\eps}, sometimes
$>10^{40}$ {\eps}) considerably exceed the Eddington luminosity of a neutron
star (e.g., Fabbiano 1996; Makishima et al. 2000 and references
therein). An ULX may be a stellar mass black hole radiating near or
above the Eddington luminosity (Begelman 2002), a stellar mass black
hole or neutron star with beamed X-ray emission (Reynolds et al. 1997; 
King et al. 2001;
K\"ording, Falcke, \& Markoff 2002), or an intermediate mass
($\sim100M_{\odot}$ or higher, but smaller than those in AGNs) black
hole radiating near or lower than the Eddington luminosity (Matsumoto
et al. 2001, Kaaret et al. 2001).

Since the launch of the {\chandra} X-ray observatory, which has
unprecedented spatial resolution ($\approx0.5$ arcsec) as well as
moderate spectral resolution over a broad band (0.1--10 keV), X-ray
source populations and luminosity functions in nearby galaxies have
been extensively studied.  One pertinent finding is that starburst
galaxies often contain ULXs (e.g., Kilgard et al. 2002 for a mini
survey; M82: Matsumoto et al. 2001, Kaaret et al. 2001, Griffiths et
al. 2000; the Antennae [NGC 4038/39]: Fabbiano, Zezas, \& Murray
2001; the Circinus galaxy: Bauer et al. 2001, Smith \& Wilson 2001;
NGC 3256: Lira et al. 2002; NGC 3628: Strickland et al. 2001).  This
finding strongly suggests that many ULXs are associated with ongoing
star formation, and perhaps with massive stars.

  Studies of spectra and variability of ULXs are essential to
understand their nature. Detection of time variability provides strong
support for the hypothesis that the observed luminosity originates in
a single source, not multiple sources. Investigations of spectral
variability are important to pursue the origin of the X-ray
emission. Flux and spectral variability has been reported for 
several ULXs (e.g., the Antennae, Fabbiano et al. 2003).
For example, a soft/hard spectral transition has been observed
in the two ULXs in the nearby galaxy IC 342 (Kubota et
al. 2001a). Such behavior is reminiscent of Galactic black hole
candidates and strongly favors an accreting black hole origin for
these sources.
If a black hole plus accretion disk interpretation of
ULXs is correct, the variability of the source luminosity and the
temperature and inner radius of the disk can be used to constrain the
structure of an accretion disk radiating at near or above the
Eddington rate (Mizuno et al. 2001; Watarai, Mizuno, \& Mineshige
2001). 
If an accretion rate is high and a disk is geometrically thick,
collimation (non-relativistic mild beaming) might be important
(e.g., King et al. 2001, King 2002).
Alternatively, the X-ray emission could originate in a disk
corona or from a relativistic jet. The luminosity function of the
X-ray sources and the environment in which the ULXs reside are also
important clues to their origin (e.g., Zezas \& Fabbiano 2002).

  The nearby (8.4 Mpc; Feldmeier et al. 1997) spiral galaxy NGC 5194
and its companion NGC 5195 (together comprising M51) represent one of
the best targets for studying a galactic X-ray source population,
including ULXs. Previous observations with the {\rosat} PSPC and HRI
have revealed the presence of luminous X-ray sources, including 2--4
sources whose luminosities exceed $10^{39}$ {\eps} (Marston et
al. 1995; Ehle, Pietsch, \& Beck 1995; Read, Ponman, \& Strickland
1997; Roberts \& Warwick 2000; Colbert \& Ptak 2002).  Additionally,
since the galaxy is near face-on, absorption by gas in the galaxy disk
is much less serious than in highly inclined systems.  Furthermore,
M51 fits nicely onto a single {\chandra} CCD, greatly simplifying the
analysis of the X-ray source population.

  In this paper, we present the properties of X-ray sources in NGC
5194 and NGC 5195 obtained with two {\chandra} observations separated
by almost exactly a year.  A {\chandra} study of the Seyfert 2 nucleus
of NGC 5194 and its extended bi-polar X-ray lobes can be found in
Terashima \& Wilson (2001). For an assumed distance of 8.4 Mpc, 1
arcsec corresponds to 40.7 pc.

This paper is organized as follows. In section 2, observations and
data reduction are presented.  Section 3 is devoted to the source
detection procedure, the source list and locations, variability,
spectra, and the X-ray luminosity function. A discussion of the nature
of the X-ray sources is given in section 4, and section 5 summarizes
the results and conclusions.

\section{Observation and Data Reduction}

{\chandra} observations of M51 were performed on 2000 June 20 and 2001
June 23 with the ACIS instrument. The background was stable in both
observations and effective exposure times of 14.9 ksec and 26.8 ksec
were obtained for the observations in 2000 and 2001, respectively.
The two observations were done with almost identical pointing.  The
offsets between the two pointing positions and roll angles were
$\approx$2 arcsec and 2\fdg3, respectively. The nucleus of NGC 5194
was placed at the aim point on the back-illuminated CCD S3 chip and
almost all of NGC 5194 and NGC 5195 fall on this chip. In this paper,
we use the S3 chip only. The data were processed by CIAO version 2.1
and CALDB version 2.7. Only {\asca} (``good'') grades 0, 2, 3, 4, 6
were used in the analysis.

\section{Results}

\subsection{X-ray Image and Source Detection}

A color image adaptively smoothed with the CIAO task ``csmooth'' is
shown in Fig. 1. The three energy bands 0.3--1.5 keV, 1.5--3 keV, and
3--8 keV correspond to red, green, and blue, respectively.  In this
image, we see extended emission distributed over NGC 5194, a bright
nuclear region (see also Terashima \& Wilson 2001), and many discrete
sources.  The companion galaxy NGC 5195 also shows discrete sources
and diffuse emission. The color indicates that the diffuse emission
has a very soft spectrum, while the discrete sources show a range of
spectral hardness.  The spectra of the discrete sources are presented
in section 3.4.

We used the ``wavdetect'' program in the CIAO package for detecting
sources. The wavelet source detection threshold was set to be
$10^{-6}$, which implies one or less false source would be detected on
the entire S3 chip. Wavelet scales of 1, $\sqrt 2$, 2, $2\sqrt 2$, 4,
$4\sqrt 2$, 8, $8\sqrt 2$, and 16 pixels were used (the pixel size of
the ACIS corresponds to 0.49 arcsec). We ran wavdetect for three data
sets: the observation in 2000, that in 2001, and the combined data of
the two. The source searches were performed in the three energy bands
0.5--8 keV, 0.5--2 keV, and 2--8 keV. We refer to these three bands as
full, soft, and hard, respectively.  We thus obtained nine source
lists, which were combined into one final source list. The source
candidates in each list were carefully examined in the raw images by
eye and we decided to include only sources with a significance greater
than 3$\sigma$ in our source list and analysis.  The nuclear region of
NGC 5194, containing the Seyfert nucleus and associated outflows, is
very bright in X-rays (see Terashima \& Wilson 2001), so emission from
within the rectangular region $\alpha = 13^{\rm h} 29^{\rm m} 51\fs9 -
53\fs4$, $\delta = 47^{\circ} 11^{\prime} 36^{\prime\prime} -
56^{\prime\prime}$ (J2000) is excluded from all the tables presented
in this paper.  The counts obtained from ``wavdetect'' were compared
with the results of manual photometry, in which extraction radii of
4$-$10 pixels were used, depending on the off axis angle of the
source.  Background counts were determined from the region surrounding
the source, and nearby sources were excluded. The number of counts
obtained from each source by manual photometry was usually in good
agreement with that obtained by wavdetect.  For a few sources, the
measured counts disagreed with the number obtained by wavdetect. In
such cases, we examined the images and decided to use the results of
the manual photometry.

Some sources were detected in only one or two out of the three energy
bands.  We calculated upper limits to the counts in the undetected
band(s) at 95\% confidence level by interpolating between the values
in Table 2 of Kraft, Burrows, \& Nousek (1991). A source detected in
only one or two out of the three data sets (observations 1, 2, and 1+2
combined) was treated in a similar way.  Upper limits to the counts in
the source region and counts in the surrounding background region were
determined by manual photometry using the same procedure as used
above.

The numbers of sources detected in each energy band and each data set
after this screening are summarized in Table 1. 113 sources are
detected on the entire S3 chip.  The source locations were compared
with an optical image (Thilker et al. 2000).  Sources within
$200^{\prime\prime}$ of the nucleus of NGC 5194 are assumed coincident
with NGC 5194 and those within $75^{\prime\prime}$ of the nucleus of
NGC 5195 are taken to be coincident with NGC 5195 (Tables 2 and 3,
respectively). Other sources detected on the S3 chip are shown in
Table 4.  In these tables we give: the source identification number
used in this paper (column 1), right ascension (column 2), declination
(column 3), {\chandra} source name (column 4), observed counts in the
full, soft, and hard bands for the combined data set of the first and
second observations (columns 5--7), observed counts in the first
observation (columns 8--10), and observed counts in the second
observation (columns 11--13). The number of counts in these tables is
not corrected for vignetting and the quoted errors on the counts
represent the $1\sigma$ range. The brightest source is
CXOM51~J133007.6+471106 in the first observation (0.055 counts
s$^{-1}$ in the 0.5--8 keV band). The pile up fraction is a few
percent for this source. Therefore, pile up is negligible for all
sources.

The X-ray position of the Seyfert nucleus ($13^{\rm h}29^{\rm
m}52\fs72, 47^{\circ}11^{\prime}42\farcs7$ (J2000)) is in good
agreement with the position of the radio nucleus ($13^{\rm h}29^{\rm
m}52\fs7101\pm0\fs0016, 47^{\circ}11^{\prime}42\farcs696\pm0\farcs026$
(J2000)) measured with the VLA in A-configuration at 8.4 GHz,
Kaiser et al.  2001).  The source CXOM51 J132938.6+471336 is also in
agreement with the position of a star. Therefore, we did not make any
corrections to the source positions.

The detection limit is about seven counts in each observation, which
corresponds to 3.9$\times10^{-15}$, 2.2$\times10^{-15}$, and
1.4$\times10^{-15}$ {\eps}{\pcm}, for the first observation, the
second observation, and the sum of the first and second observations,
respectively, in the 0.5--8 keV band for a typical spectrum (a power
law with a photon index of 1.5 absorbed by the Galactic column {\NH} =
$1.3\times10^{20}$ {\pcm} (Stark et al. 1992)). According to the $\log
N - \log S$ relation from the {\chandra} deep surveys (Mushotzky et
al. 2000; Tozzi et al. 2001; Brandt et al. 2001), the expected number
of background sources is 8--12 on the entire S3 chip above the flux
limit for the sum of the two observations, where the effects of
vignetting and bad pixels have been taken into account to calculate
our survey area.  Since the detected number (17) of sources listed as
outside NGC 5194 and NGC 5195 (Table 4) is larger than that expected
for background sources (4--6), some of the sources shown in Table 4
are apparently associated with M51.  

% * A fraction of these sources might be globular clusters (GCs)
% containing an X-ray source. Optical identifications of X-ray sources
% in external galaxies have shown majority of bright X-ray sources with
% $> 5-10\times10^{38}$ {\eps} are associated with GCs in at least
% several galaxies studied so far (Di Stefano et al. 2003, Kong et
% al. 2002, Kundu et al. 2002, Angelini et al. 2001). Note also that
% only a few \% of the X-ray sources in M81 appear to be in globular
% clusters (Swartz et al. 2003).  A catalog of GCs in M51 appropriate
% for statistical studies and identifications of X-ray sources is not
% available, however.  **

\subsection{Source Locations}

\subsubsection{Comparison with {\em ROSAT} Sources}

We have compared our source list with previous X-ray observations.
The best X-ray source list before {\chandra} was obtained with
{\rosat} (Ehle, Pietsch, \& Beck 1995 (HRI); Marston et al. 1995
(PSPC); Roberts and Warwick 2000 (HRI); Read, Ponman, \& Strickland
1997 (PSPC); Colbert \& Ptak 2002 (HRI)). Correspondences between
{\chandra} and {\rosat} sources are given in Table 5, where we list
all the {\rosat} sources within 12$^{\prime\prime}$ of a given
{\chandra} source (the {\rosat} positions have uncertainties of
$\approx10^{\prime\prime}$).

Four {\rosat} sources have two {\chandra} counterparts in their error
circles (NGC 5194 {\chandra} \#5 and 6, \# 27 and 28, \#41 and 42, and
\# 69 and 70). This result demonstrates that luminous X-ray sources
found with low-spatial resolution can be superpositions of two or more
sources and that interpretation should be made with caution. One
{\rosat} source (source 7 in Ehle et al. = source 4 in Marston et
al. = source 4 in Read et al.) has no {\chandra} counterpart. The
{\chandra} counterparts of two {\rosat} sources (source 5 in Marston
et al. = source 3 in Read et al., and source 6 in Ehle et al. = source
3 in Roberts \& Warwick) are very faint (9 counts and 29 counts,
respectively, in the 0.5--8 keV band for the total of the two
{\chandra} observations). The first ({\chandra} \#16) is almost at the
detection limit of {\chandra}.  These {\rosat} sources with no or a
very faint {\chandra} counterpart, indicating drastic flux decreases
from the {\rosat} to the {\chandra} observations, are likely to be
transient sources.

The brightest source (CXOM51~J133007.6+471106 = NGC 5194\#82) may be
identified with a source detected in an {\Einstein} HRI image (Palumbo
et al. 1985) and a {\sax} image (Fukazawa et al. 2001).

\subsubsection{Optical Counterparts}

In Fig. 2, the positions of the detected sources are overlaid on a
narrow-band optical image including H$\alpha$ and the nearby continuum
(Thilker et al. 2000). Most of the discrete sources in NGC 5194 are
located in or near the spiral arms, but sources are also found between
the arms. A few X-ray sources appear to be associated with bright
\ion{H}{2} regions, while others show no clear association with
optical sources. All 7 of the ULXs in NGC 5194 are in, or close to, a
spiral arm, supporting an association between ULXs and star formation.

We cross-correlated the X-ray source positions with the list of
\ion{H}{2} regions in NGC 5194 obtained by Thilker et al. (2000).
Among the 1625 \ion{H}{2} regions in their catalog above a 5$\sigma$
detection limit of $L_{\rm H\alpha}$ = $10^{36}$ {\eps}, about 1350
fall on {\chandra} chip S3. We found that 22 of these 1350 \ion{H}{2}
regions have an X-ray source within 1.5$^{\prime\prime}$ or 61 pc.
Most of the 22 \ion{H}{2} regions are very faint, and only two bright
\ion{H}{2} regions have a candidate X-ray counter part (NGC 5194 \#55
= CXOM51~J132955.5+471402 and NGC 5194 \#59 =
CXOM51~J132955.9+471145). These two X-ray sources are faint in X-rays
(17 and 43 counts, respectively, were detected in the 0.5--8 keV band
in the sum of the two observations).

Four (NGC 5194 \#37 = CXOM51~J132953.3 +471042, NGC 5194 \#41 = CXOM51
J132953.7 +471436, NGC 5194 \#69 = CXOM51 J133001.0 +471344, NGC 5194
\#82 = CXOM51 J133007.6 +471106 ) out of the seven ULXs have an
extremely faint \ion{H}{2} region, which is not visible in Fig 2,
within 1\farcs5. One ULX (NGC 5194 \#26 = CXOM51~J132950.7+471155) is
2\farcs3 away from an \ion{H}{2} region and is in a chain of bright
\ion{H}{2} regions along an inner spiral arm.

  30 point sources have been detected in multi-band {\hst} WFPC2
images of the central ($11^{\prime \prime}\times16^{\prime \prime}$,
or 450$\times$650 pc) region (Table 2 in Lamers et al. 2002).  Lamers
et al. (2002) argue that most of these point sources are isolated
massive stars, rather than clusters.  Unfortunately, most of these
sources are located in the region of the bright, nuclear, diffuse
X-ray emission, which probably originates from gas heated by the jets
from the AGN (Terashima \& Wilson 2001), and no X-ray point sources
are seen at the positions of the {\hst} sources. A few {\hst} sources
are outside the diffuse X-ray emission but we find no X-ray
counterparts.

  Globular clusters appear to be the dominant sites of luminous low
mass X-ray binaries in early-type galaxies (Angelini et al. 2001;
Kundu et al. 2002; Di Stefano et al. 2003). However, the fraction of
the unresolved X-ray sources in the spirals M81 (Swartz et al. 2003)
and M31 (Kong et al. 2002) that are associated with known globular
clusters is small. Unfortunately, we cannot evaluate the association
between compact X-ray sources and globular clusters in M51 as we are
unaware of any catalog of globular clusters in this galaxy.

\subsubsection{Radio Counterparts}

  We searched for counterparts in a FIRST VLA survey image at 1.4 GHz
and in the source list from VLA 6 and 20 cm observations by Ho \&
Ulvestad (2001; their Table 4) and found no radio counterparts, except
for the nuclei of NGC 5194 and NGC 5195.

\subsection{Variability of Point Sources}

\subsubsection{Short-term Variability}

  We have searched for short-term variability of relatively bright
point sources by investigating their fluxes within each
observation. There are 16 and 22 sources with $>$ 70 counts per
observation in the 0.5$-$8 keV band in the first and second
observation, respectively.  Light curves with a time bin of 2000 sec
were created and examined with the Kolmogorov-Smirnov (K-S) test to
evaluate the null hypothesis that any given source has constant
flux.

%src 37, src 26

  In the first observation, NGC 5194 \#63 (= CXOM51~J132957.6+471048)
shows a low KS probability ($2\times10^{-7}$) (Fig 3a). Another object
(NGC 5194 \#69 = CXOM51~J133001.0+471344) shows two peaks separated by
$\approx$ 7000 sec (Fig. 3b), which may be a periodicity or two
physically distinct flux peaks. Liu et al. (2002) interpreted these
features as indicating a 2.1 hr period, although only two periods are
covered by the observation.

%src 17

  In the second observation, one source (NGC 5194 \#76 =
CXOM51~J133004.3+471321) clearly shows a declining flux, going nearly
to zero in the second half of the observation. The light curve of this
object is shown in Fig. 3c. The light curves of other sources have KS
probabilities greater than $1\times10^{-4}$ and variability is not
significant. NGC 5194 \#69 was very dim (46 counts detected in
0.5--8 keV) in the second observation, and the variability seen in
the first observation is not detected.

\subsubsection{Long-term Variability}

  We compared the count rates obtained in the two separate
observations as a means of searching for variability on a time scale
of a year.  Fig. 4 shows the ratio of the count rate, C(0.5--8 keV),
in the second observation to that in the first observation plotted
against the luminosity in the 0.5--8 keV band for the sources in NGC
5194 and NGC 5195.  The luminosities were calculated from the total
counts of the two observations, assuming a power-law spectrum with a
photon index derived from the band ratio, C(2--8 keV)/C(0.5--2 keV),
and absorption by only the Galactic column.  When only an upper or
lower limit on the band ratio is available, the typical photon index
of 1.5 was assumed if the band ratio is consistent with this value;
otherwise we used the limit on the band ratio to estimate the spectral
shape.  The numbers of objects in NGC 5194 which show time variability
at greater than the $3\sigma$ level in the full, soft, hard, and any
one of these three bands are shown in Table 1. About one-third of
discrete sources in NGC 5194 exhibit time variability, which strongly
implies that these sources are single (or at most a few) compact
objects, and are most likely to be black hole binaries powered by
accretion, given their large luminosities (\simgt $10^{38}$ ergs
s$^{-1}$).  Their band ratios and spectra are also consistent with an
X-ray binary interpretation (section 3.4), except for a few cases. A
small fraction could be recent supernovae, which show variability on a
timescale of $\sim$ 1 yr (e.g., Immler \& Lewin 2002).

Four objects show large amplitude variability (a factor of ten or
more), suggesting a transient nature. Two of them are ULXs,
demonstrating that there are two classes of these objects, i.e.,
transient and persistent ULXs.

\subsection{X-ray Spectra of Point Sources}

\subsubsection{Spectra of Bright Point Sources}

Spectral fits were performed for sources with $>$70 detected counts in
the 0.5--8 keV band in one observation using XSPEC version 11. In all,
spectral models were obtained for 38 spectra of 26 sources. The
spectra were grouped so that each spectral bin contained at least 15
counts, and the $\chi^2$ minimization technique was employed in the
fit unless otherwise noted. We obtained a background spectrum for each
source from an adjacent field and subtracted the background from the
source spectrum. Any sources in the background region were
excluded. Two models --- a power law and a multicolor disk blackbody
(MCD, Mitsuda et al. 1984, Makishima et al. 1986) --- were examined in
spectral fitting. In some cases, these models did not provide
satisfactory representations of the data and other models were
examined, as discussed below. The results of the spectral modeling are
summarized in Tables 6 and 7 for objects with luminosities in the
0.5$-$8 keV band greater than and less than $10^{39}$ {\eps},
respectively.  Note that a given source may have $L$(0.5 -- 8 keV) $>$
10$^{39}$ {\eps} in one observation or when modeled with a particular
choice of spectrum, and $L$(0.5 -- 8 keV) $<$ 10$^{39}$ {\eps}
otherwise.  The errors or parameter ranges shown in this paper
represent the 90\% confidence level for one parameter of interest
($\Delta \chi^2=2.7$) unless otherwise stated. Luminosities for a MCD
model are calculated by assuming the inclination of the disk
$i=60^{\circ}$, i.e., $L = 2\pi D^2 F (\cos i)^{-1} = 4\pi D^2 F$,
where $F$ is the observed flux and $D$ is the distance. Errors,
fluxes, and luminosities are not shown for unacceptable fits.

\subsubsection{Band ratios}

Figs. 5a and 5b show the dependence of the band ratio on counts in the
0.5--8 keV band for sources in NGC 5194 and NGC 5195, respectively;
the summed data of the two observations are used. The counts are as
observed and not corrected for the effects of vignetting and the edge
of the CCD chip. The band ratios for power law spectra modified by
Galactic absorption are shown as dashed lines for photon indices of
--1.0, 0.0, 1.0, 2.0, and 3.0. We calculated the source luminosities
as described in section 3.3.2. The dependence of the band ratio on
luminosity is displayed in Figs. 6a and 6b for NGC 5194 and NGC 5195,
respectively.  These plots show that most of the relatively luminous
($>10^{38}$ {\eps}) sources have spectra consistent with a power law
with a photon index between 1 and 2, while some sources have very flat
or steep spectra. There is no obvious dependence of photon index on
luminosity.

\subsection{Luminosity Function}

Using the luminosities derived earlier (from the total counts in the
two observations - section 3.3.2), we have calculated the cumulative
luminosity function in the 0.5 - 8 keV band for the X-ray sources in
NGC 5194 (Fig. 7).  The luminosity function was modelled as a power
law function above $10^{38}$ ergs s$^{-1}$, where it is virtually free
from incompleteness. We confirmed that incompleteness is negligible
above $10^{38}$ ergs s$^{-1}$ by comparing our source and background
counts with the simulations presented in Kim \& Fabbiano (2003). We
used a maximum-likelihood method and calculated errors by means of the
Gehrels approximation (Gehrels 1986). A power-law slope of $\alpha
=0.91\pm0.14$ ($N(>L$(0.5 -- 8 keV)) $\propto$ $L$(0.5 -- 8
keV)$^{-\alpha}$) was obtained. The luminosity function does not
require a break around the Eddington luminosity for $1.4M_{\odot}$,
as has been found in several early type galaxies (Sarazin,
Irwin, \& Bregman 2000; Blanton, Sarzin, \& Irwin 2001; Irwin,
Sarazin, \& Bregman 2002).

In the model fit, the errors in the number of sources in each bin were
taken into account while other sources of error, such as uncertainties
in photon count rates and spectral shapes, were ignored.  These latter
effects are, however, relatively small since we discarded the
low luminosity ($<10^{38}$ {\eps}) sources, of which there
are many but the photon statistics for each source are poor
(e.g., Zezas \& Fabbiano 2002).

\section{Discussion}

\subsection{Incidence of Ultra Luminous X-ray Sources and Luminosity Function
of X-ray Sources}

Our observations have detected many luminous sources, among which 7
and 28 sources in NGC 5194 have a luminosity (0.5--8 keV) greater than
$10^{39}$ and $10^{38}$ {\eps}, respectively, assuming they are
associated with M51. Two source in NGC 5195 has a luminosity in excess
of $10^{39}$ {\eps}. The number of ULXs, defined as $L$(0.5--8
keV)$>10^{39}$ {\eps}, is much higher than typically seen in spiral
galaxies (0--2; e.g., Roberts \& Warwick 2000). The large number of
luminous X-ray sources is not simply related to the number of young
stars in M51.  In fact, the absolute B magnitude, which is often used
as a measure of the relatively young stellar population, of NGC 5194
is modest ($M^0_{B_T}=-20.95$, calculated from information in the RC3
catalog for our assumed distance) in comparison with the other spiral
galaxies studied by Roberts \& Warwick (2000).  Indeed, the majority
of galaxies with similar values of $M_{\rm B}$ to M51 possess no ULXs.

NGC 5194 is a classical, nearby Sc galaxy.  It thus has star formation
associated with density-wave driven spiral arms in an ISM rich in
molecular gas (e.g., Scoville et al. 2001).  Most of the detected
X-ray sources are located in or near the spiral arms
(Fig. 2). Further, all 7 of the ULXs in NGC 5194 are in, or very close
to, a spiral arm.  Our results thus confirm that the ULXs in M51 are
associated with star formation, but in a different context to
starburst galaxies, such as NGC 1068 (a very luminous circumnuclear
disk starburst possibly related to the AGN; see Smith \& Wilson [2003]
for the X-ray source population) and interacting galaxies, such as the
Antennae (galaxy merger-driven star formation; see Zezas \& Fabbiano
[2002] for the X-ray source population).

As noted in section 3.5, the cumulative luminosity function for the
X-ray sources in NGC 5194 is well fitted by a power law with a slope
of 0.91 above $10^{38}$ {\eps} (Fig. 7). This slope is inbetween the
values found in disk galaxies (1.1--1.4) and active starforming
galaxies (0.5--0.8 for M82, NGC 253, and the Antennae), and
significantly lower than those derived for ellipticals and lenticulars
(Kilgard et al. 2002 and references therein). A break in the
luminosity function (i.e. a change in its slope) can potentially be
used to measure the formation history and age of the X-ray source
population (Kilgard et al. 2002, Wu et al. 2003) and the amount of
beaming in luminous sources (K\"ording et al. 2002). Such a break is
not apparent in the luminosity function of NGC 5194. A power-law
shape without a break is in agreement with luminosity functions
observed for X-ray sources in starburst galaxies and in the galaxy disk
of M81 (Tennant et al. 2001, Swartz et al. 2003), and is indicative of
on-going star formation (Wu et al. 2003).

\subsection{Spectra and Variability of Luminous and Ultra-luminous X-ray 
Sources}

We have modelled the X-ray spectra of bright sources with two
continuum models -- a power law and a MCD (Tables 6 and 7). In this
subsection, we discuss interpretations of the spectra and their
variability in ULXs.

\subsubsection{ULXs with a Power Law Spectrum}

%--- power law spec: src9 & src 67

The spectra of NGC 5194 \#37 = CXOM51~J132953.3+471042 and NGC 5194
\#82 = CXOM51 J133007.6+471106 (Fig. 8) in the second observation
(June 2001) are better fitted with a power law than a MCD model (Table
6). The {\asca} spectra of ULXs studied by Makishima et al. (2000) are
better modeled by a MCD than a power law. However, ULXs whose spectra
are best fitted by a power law have been found in recent observations
(e.g., a source in Holmberg II observed with {\asca} [Miyaji, Lehmann,
\& Hasinger 2001]; sources in IC 342 observed with {\asca} [Kubota et
al. 2001a]; a source in NGC 3628 observed with {\chandra} [Strickland
et al. 2001]; a source in NGC 4565 observed with {\xmm} (RXJ
1236.2+2558 = XMM J123617.5+285855 = CXO J123617.4+255856 [Foschini et
al. 2002]; NGC 5204 X-1 [Roberts et al. 2001]; several of the luminous
sources in NGC 1068 observed with {\chandra} [Smith \& Wilson
2003]). Thus it appears that the two sources in NGC 5194, along with
these objects, constitute a class of ULXs whose spectrum is best
described by a power law.

A power law spectrum may arise in several scenarios.
Phenomenologically, a power-law like spectrum is found in the low/hard
state or very high state of a black hole binary (e.g., Tanaka 1996,
Done 2002). Physically, power law spectra may result from
non-thermal, possibly relativistically beamed, X-ray emission from
jets, as in blazars (K\"ording, Falcke, \& Markoff 2002), or through
Comptonization of photons from the accretion disk, as is believed
to occur in the low/hard state and the very high state of a binary
containing a black hole (Done 2002; Kubota, Done, \& Makishima 2002).

If the power-law spectra are analogous to the low/hard state in
Galactic black hole candidates and the state transition is governed by
the mass accretion rate, the ULXs would be emitting at less than a few
\% of $L_{\rm Edd}$ (e.g., Done 2002). If this is the case, a very
high mass ($>$ a few times 100 $M_{\odot}$) black hole is required.
It is, however, not clear that such an extrapolation from stellar mass
to such high mass black holes is valid.

On the other hand, if the observed X-ray emission is significantly
beamed toward us, the intrinsic luminosity would be much lower than we
have derived by assuming isotropic radiation. In this case, basic
quantities such as the mass of the compact object, mass accretion
rate, and beaming factor cannot be obtained from the present data.

Alternatively, spectra which are good approximations to power law
shapes may be the results of Comptonization of thermal emission from
an accretion disk with a high accretion rate. Such strong
Comptonization is believed to occur in Galactic jet sources (e.g.,
GRO~J1655-40, Kubota, Makishima \& Ebisawa 2001b ) and possibly an
ULX in IC 342 (Kubota et al. 2002). Kubota et al. (2002) reanalyzed
the {\asca} spectrum of source 1 in IC 342 observed in 2000. The
spectrum can be fitted with either a power law ($\Gamma =
1.73\pm0.06$, modified by an ionized Fe-K edge at $8.4\pm0.3$ keV) or
a Comptonized blackbody with an inner disk temperature of $kT_{\rm
in}=1.1\pm0.3$ keV. Kubota et al. (2002) argued that the latter model
is a more appropriate interpretation of the apparent power law shape
observed in this source based on their detailed spectral analysis and
the analogy with the spectral variability and Comptonization in
Galactic jet sources. This type of spectral state was observed in
GRO~J1655--40 and XTE~J1550--564 when the bolometric luminosity of the
disk was around $\sim$10-20\% of its Eddington luminosity (Kubota et
al. 2001b; Kubota 2001). In their analysis, the bolometric luminosity
is calculated from the MCD parameters of the underlying disk emission
(see section 4.2.5). If the bolometric luminosities of the two sources
in NGC 5194 with power law spectra correspond to this Eddington ratio
($L_{\rm bol}/L_{\rm Edd}=0.1$), the masses of the compact object are
$>$ 98 $M_{\odot}$ (source 37) and $>$ 120 $M_{\odot}$ (source 82 in
2001) taking $L_{\rm Edd}$ = 1.5 $\times$ 10$^{38}$ (M/M$_{\odot}$)
ergs s$^{-1}$ (assuming spherical symmetry and a hydrogen to helium
composition ratio of 0.76:0.23 by weight [e.g., Makishima et
al. 2000]). However, it is not clear that such spectral behavior {\it
always} appears at such an Eddington ratio. If the power-law state
appears between the optically thick geometrically thin disk and the
slim disk, $L_{\rm bol}/L_{\rm Edd}$ would be 1--2. In this case, the
masses are estimated to be $>$ 5 $M_{\odot}$ (source 37) and $>$ 6
$M_{\odot}$ (source 82 in 2001).

NGC 5194 \#37 is detected in only the second observation, indicating
the photon flux increased by a factor of 100 in the 0.5$-$8 keV
band. Although this large variability may imply that the source is a
transient X-ray binary, such behavior is also consistent with any of the
possibilities described above.

On the other hand, NGC 5194 \#82 is detected, and spectral information
is available, in both observations. The spectrum flattened ($\Gamma =
2.26$ to 1.86) and the flux decreased (by 40\%) from the first to the
second observation.  This trend is consistent with that of Galactic
BHCs in the low/hard state and in the state with strong
Comptonization.  In both states, steeper spectra are seen in higher
flux states.

Thus, no strong constraints on the masses of the compact objects and
mass accretion rates in these ULXs can be obtained. Future
observations of spectral variability should be able to quantify the
changes in spectral state, and hopefully allow fundamental
parameters, such as black hole mass and mass accretion rate, to be
determined.

% --- Transient ULXs src 26, (the transient src 67 already discussed)

\subsubsection{Spectral Steepening in NGC 5194 \#69}

NGC 5194 \#69 = CXOM51~J133001.0+471344 shows remarkable spectral
variability (Fig. 9). The photon flux in the 2--8 keV band decreased
by a factor of more than 13 from the year 2000 to the year 2001
observation. The first spectrum is relatively hard: $\Gamma =
1.24^{+0.12}_{-0.17}$ (power law) or $kT_{\rm in}=2.3^{+1.0}_{-0.5}$
keV (MCD), while the second spectrum is extremely soft: $\Gamma>5.1$
(power law), $kT_{\rm in}=0.17^{+0.13}_{-0.06}$ keV (MCD), or $kT =
0.16^{+0.08}_{-0.06}$ keV (black body) (Table 6).  The absorbing
column increased from $3.4^{+3.9}_{-3.4}\times10^{20}$ {\pcm} to
$8.2^{+8.0}_{-5.0}\times10^{21}$ {\pcm} from year 2000 to 2001 (values
are for MCD models --- Table 6). Our results are in general agreement
with those of Liu et al. (2002), who analyzed the same observations.
If we adopt a power law model and a MCD model for the first and second
spectra, respectively, the intrinsic luminosity of this source
declined by a factor of 3460 in the 2--10 keV band (the precise
factor depends on the choice of spectral model). This object is the
first example of such a drastic spectral change in ULXs, and is
potentially extremely important for elucidating their true nature.

Such drastic steepening of the spectrum accompanied by a large flux
decline is seen in soft X-ray transients. Soft X-ray transients show
very soft spectra ($kT=0.2-0.3$ keV for a black body model) when the
luminosity goes below $10^{34}$ {\eps} (Tanaka 1996, Tanaka \&
Shibazaki 1996). Such luminosities, however, are much lower than we
observe in source NGC 5194 \#69 in the second observation
($5.6\times10^{38}$ {\eps} in 0.5--8 keV and $7.4\times10^{35}$ {\eps}
in 2--10 keV, the latter measured by extrapolating the best-fit MCD
model up to 10 keV). For $L$(2--10 keV) $\approx 6\times10^{35}$
{\eps}, Galactic black hole candidates are in a low/hard state, with
spectra described by a power law with a photon index of 1.4--1.9,
quite different to source \#69 in 2001. Thus, it is not clear whether
the similarity between the spectral shapes of the quiescent state of
soft X-ray transients and NGC 5194 source \# 69 in the second
observation has any physical significance. If the transition of the
spectra is governed by the mass accretion rate, observations of a
state transition can be used as a measure of the mass of the central
object and the mass accretion rate. The luminosity of NGC 5194 \#69 in
the state showing the very soft spectrum is more than 70 times higher
than that of the similar state in stellar-mass black holes. If a
simple scaling with mass is applied, the mass of the compact object in
NGC 5194 \#69 is $\simgt$ 70 times larger, i.e.  $\simgt$ several
times 100 $M_\odot$, which is in the range of intermediate-mass black
holes (IMBHs). An improved understanding of the behavior of soft X-ray
transients in our Galaxy, and monitoring of extragalactic luminous
X-ray sources, will be crucial to pursue the true nature of luminous
transient sources.  Future observations of NGC 5194 \# 69 are needed
to confirm the possible periodicity of 2.1 hrs (see Fig. 3b); possible
interpretations were discussed by Liu et al. (2002).

\subsubsection{Emission Lines in CXOM51~J 132950.7+471155 = NGC 5194 \#26}

% src79

  CXOM51~J 132950.7+471155 = NGC 5194 \#26 has the largest band ratio of
all detected sources in M51 (Figs. 5a and 6a). This source can be seen
as a blue (= very hard) source to the NW of the nucleus in the color
image shown as Fig. 1 in Terashima \& Wilson (2001).

  This source shows a remarkable spectrum and variability. As shown
below, the spectrum includes emission lines, and we use a spectrum
with a finer-bin size to bring out these features. In
the spectral fits, we adopted a maximum likelihood method using the
C-statistic (Cash 1979), instead of $\chi^2$ minimization, since
each bin contains a small number ($>3$) of counts.

In the first observation (June 2000, Fig. 10a), the spectrum is quite
hard and virtually no photons are detected in the low energy region
below 1.5 keV. Since there are indications of emission lines, we added
Gaussians to an absorbed power law continuum to model the
spectrum. The results are summarized in Table 8, which includes
identifications for the lines, assuming they are K$\alpha$
transitions. The intrinsic luminosity corrected for absorption is
$3.4\times10^{39}$ {\eps} in the 0.5$-$8 keV band.  We examined the
statistical significance of these emission lines by Monte Carlo
simulations. We simulated 2000 spectra with the best-fit continuum
model and no emission lines, and fitted them with a model consisting
of an absorbed power law and a Gaussian line. The line center energy
was fixed at one of the four energies (1.8, 3.2, 4.0, and 6.7 keV)
reported in Table 8. The width was fixed at 0. We calculated the
probability that a fake line stronger than the actual data (Table 8)
would be detected and obtained 4.0, 0.10, 0.15, and 0.40\% for the
lines at 1.8, 3.2, 4.0, and 6.7 keV, respectively. Thus the detection
of these lines is highly significant.

On the other hand, emission-line features are not clearly seen in the
spectrum obtained in the second observation (June 2001, Fig. 10b). In
this case, the continuum is modeled as a partially-covered power law,
and the spectral parameters are shown in Table 8. The intrinsic
luminosity in the second observation ($5.0\times10^{39}$ {\eps} in the
0.5--8 keV band) is 50\% higher than that in the first observation.

Emission lines from highly ionized atoms and with large equivalent
widths are observed in eclipses of high-mass X-ray binaries, when the
direct emission from the compact object is blocked by the companion
star, and only emission from and scattered by the photoionized medium
surrounding the system (such as a stellar wind from the companion) is
observed (e.g., Nagase et al. 1994; Ebisawa et al. 1996; Sako et
al. 1999; Schulz et al. 2002).  In such a situation, the intrinsic
luminosity would be much larger (typically by a factor of 10$-$100;
e.g., Nagase et al. 1994; Ebisawa et al. 1996) than the observed
luminosity. If so, the observed luminosity implies that the intrinsic
luminosity is at least $\sim 5\times10^{40}$ {\eps} if the emission is
isotropic. If this is the case, this system would be an eclipsing
ULX. The spectrum in the second observation no longer shows strong
emission lines, which implies that the contribution of direct emission
from the compact object is significant. The observed luminosity in the
second observation ($5.0\times10^{39}$ {\eps} in 0.5--8 keV), however,
is much lower than the expectation from the scattering picture
($>5\times10^{40}$ {\eps}). Therefore, we conclude that an eclipse
plus scattering scenario, with an isotropic X-ray source, is unlikely
to account for the spectrum observed in June 2000.

The above luminosity problem could be resolved if the irradiating
emission is anisotropic and beamed towards the scattering
medium.

\subsubsection{Spectra and Variability of Other ULXs}

% src 103, src 63

The spectra of the two ULXs NGC 5194 \#5 = CXOM51~J132939.5+471244
(Fig. 11a) and NGC 5194 \#41 = CXOM51~J132953.7+471436 (Fig. 11b) can
each be represented by either a power law or a MCD model (Table
6). The luminosity of the former source decreased by a factor of
$\approx$2.5, while that of the latter remained the same between the
two observations. Although the best-fit spectral parameters of both
these sources suggest that both spectra flattened, the errors are
large and variability of the spectral shape is not statistically
significant. The spectra of the source NGC 5195 \#12 =
CXOM51 J133006.0+471542 may also be described by either a power law or
a MCD model; there is no evidence for variability.

\subsubsection{Multi Color Disk Blackbody Interpretation of the Spectra}

%Sources with $10^{38} < L_{\rm X} < 10^{39}$ {\eps}

The X-ray sources with $10^{38}$ $<$ $L$(0.5 -- 8 keV) $<$ $10^{39}$
{\eps} are of significant interest since they could be a population
intermediate between ULXs and ordinary X-ray binaries.  The spectra of
most of the sources in this luminosity range for which we performed
spectral fitting (Table 7) may be described by either a power law or a
MCD model.

If a power law is the correct description of the spectra of the two
ULXs discussed in section 4.2.1 and the objects with $L$(0.5 -- 8 keV)
$<$ 10$^{39}$ {\eps}, the same possibilities as described in section
4.2.1 are viable.  Here, we examine the implications of an alternative
interpretation of the spectra --- MCD emission arising from optically
thick accretion disks.

According to the MCD formalism of the spectrum from an optically-thick
geometrically-thin accretion disk, the inner temperature of the disk
$kT_{\rm in}$ depends on the mass accretion rate and black hole mass
as $kT_{\rm in} \propto \eta^{1/4} M^{-1/4}$ (e.g., eq. (12) in
Makishima et al. 2000), where $\eta$ is the Eddington ratio ($L_{\rm
bol}/L_{\rm Edd}$) and $M$ is the black hole mass. Fig. 12 shows the
relation between the bolometric luminosity ($L_{\rm bol}$) and inner
disk temperature $kT_{\rm in}$.  The expected relationships between
$L_{\rm bol}$ and $kT_{\rm in}$ for a thin accretion disk with
fixed black hole masses ($M$ = 3, 6, 12, and 24 $M_\odot$) and fixed
Eddington ratios ($\eta$ = 0.01, 0.1, 1, and 10) are shown as
dot-dashed and dashed lines, respectively.  We have assumed that the
disk inner radius, $R_{\rm in}$, is equal to three Schwarzschild radii
($R_{\rm S}$), the ratio of the color temperature to the effective
temperature (``spectral hardening factor'') $\kappa$ = 1.7 (Shimura \&
Takahara 1995), and the factor $\xi$, which corrects for the fact that
$T_{\rm in}$ occurs at a radius somewhat larger than $R_{\rm in}$, is
$\xi$ = 0.41 (Kubota et al. 1998).  The bolometric luminosity is
calculated from $L_{\rm bol} = 4\pi (R_{\rm in}/\xi )^2 \sigma (T_{\rm
in}/\kappa)^4$. This plot is the same diagram as Fig. 3 in Makishima
et al. (2000).

For each source, we calculated $L_{\rm bol}$ from the MCD model.
The majority of the sources with $L_{\rm bol}$ {\simgt} $10^{39}$
{\eps} have the best-fit $kT_{\rm in}$ in the range 1.6--2.3 keV which
is similar to the ULXs studied by Makishima et al. (2000) and higher
than Galactic black hole candidates. The objects with $L_{\rm bol} =
(5-10)\times 10^{38}$ {\eps} are clustered around $\eta$ = 1. In other
words, their temperature is comparable to that expected from the black
hole mass determined by assuming that $L_{\rm bol}\approx
L_{\rm Edd}$ and their inferred masses are in the range of
stellar-mass black holes ($\sim$ 10 $M_{\odot}$ or less). Three sources
have relatively low temperatures ($kT\approx 0.5$ keV) and small
values of $\eta$ ($<$0.1).  The data suggest a correlation between
$L_{\rm bol}$ and $kT_{\rm in}$. 

We note that X-ray spectra of Galactic BHCs in the high/soft state
show a power law component in addition to the MCD component. If this
power law component exists in the luminous sources, a MCD model
might be too simplistic and lead to unphysical parameters. In BHCs, the
MCD component dominates the X-ray spectrum below $\sim 3$ keV, while
the power law component is significant above several keV (e.g, Dotani
et al.  1997), where the effective area of {\chandra}
decreases. Therefore, the spectra obtained in the present work 
may not be 
sensitive to a power law component that dominates at high energies. Note
also that a low-temperature disk and a significant power law component
are detected in ULXs in NGC 1313 (Miller et al. 2003), for which
Makishima et al. (2000) derived high $kT$ based on a single MCD model
fit to {\asca} spectra.

The two objects NGC 5194 \#69 = CXOM51 J133001.0+471344 (in 2000;
$kT$=2.3 keV, $\log L_{\rm bol}$ = 39.49 {\eps}) and NGC 5194 \#41 =
CXOM51 J132953.7+471436 (in 2001; $kT$ = 2.2 keV, $\log L_{\rm bol}$ =
39.11 {\eps}), and, marginally, one more object NGC 5195 \#12 = CXOM51
J133006.0 +471542 (in 2001; $kT$ = 1.8 keV, $\log L_{\rm bol}$ = 39.01
{\eps}), are not compatible with $\eta<1$.  They have too high a
temperature compared to that expected in the MCD model from the black
hole mass implied by assuming $\eta \le 1$.  This is the same as the
``too-hot-accretion-disk problem'' in ULXs (Makishima et
al. 2000). This problem may be resolved if the accretion is mildly
super Eddington, when a ``slim disk'' is expected (e.g., Watarai et
al. 2000), or the black hole is rotating and we view the system near
edge-on (Makishima et al. 2000; Ebisawa et al. 2001b; Mizuno et
al. 2001; Ebisawa et al. 2003).

The observed luminosities and temperatures of a large fraction of the
ultraluminous / luminous X-ray sources ($L_{\rm bol}$ {\simgt}
$3\times10^{38}$ {\eps}) imply relatively large mass accretion rates
($0.3 < \eta < 6$), for which an optically-thick, thin accretion disk
may not be an appropriate description. As noted above, under high
accretion rates, the accretion disk may be in a slim disk branch.  Fig
13 shows a comparison between MCD parameters inferred from our
observations and predictions of a slim disk model in the $L_{\rm
bol}$-$R_{\rm in}$ (Fig 13a) and $kT_{\rm in}$-$R_{\rm in}$ (Fig 13b)
planes (Watarai et al. 2001, errors in the original calculations have
been corrected in Fig 13). In the model, $kT_{\rm in}$ and $R_{\rm
in}$ are determined by fitting a MCD model to calculated slim disk
spectra. The parameters in these diagrams are the mass accretion rate
$\dot{m}=\dot{M}/(L_{\rm Edd}/c^2)$ and the black hole mass
$m=M/M_{\odot}$. Most of the observed data points fall to the
lower-mass side of the constant mass line $m$=10 in these figures and
are in good agreement with the slim disk prediction for a mass of
$5-10M_{\odot}$ and an accretion rate $\dot{m}$=3--100.  An
interesting feature is that the data points distribute roughly along a
constant mass line. This may indicate that the observed variety of
values of $L_{\rm bol}$ and $kT_{\rm in}$ is mainly a result of
differences in the mass accretion rate and that the black hole masses
are in a relatively small range ($5-10M_{\odot}$). Note, however, that
the calculations by Watarai et al. do not include general relativistic
effects and Compton scattering inside the disk, and that more
sophisticated models would be necessary for quantitative arguments.

Although a study of spectral variability provides a good test of a
slim disk scenario (Mizuno et al. 2001, Watarai et al. 2001), the
errors in the obtained parameters, $L_{\rm bol}$, $R_{\rm in}$, and
$kT_{\rm in}$, are too large to distinguish a slim disk from a thin
disk model. 

Mild beaming, resulting from a collimation by the inner part of the accretion
disk, has been proposed as an interpretation of the apparently super
Eddington luminosities of ULXs (King et al. 2001).  If this is the
case, objects with high $kT_{\rm in}$ and $L_{\rm bol}$ become
consistent with $\eta < 1$, though it is not clear whether such a disk
geometry is achieved for these parameters. 
King et al. (2001) envisage anisotropic X-ray emission from either
thermal emission from a radiation pressure supported torus (e.g.,
Jaroszynski, Abramowicz \& Paczynski 1980) or emission from a jet, however
produced. Thick disks are subject to instabilities and the very existence
of radiation tori is questionable. These and other issues prevent a
conclusion on the relevance of beaming to ULXs.

The objects with a lower luminosity ($L_{\rm X} < 5\times10^{38}$
{\eps}) and $kT_{\rm in}\sim 1$ keV are consistent with a thin
disk picture with a stellar-mass black hole ({\simlt} $10M_\odot$)
radiating at a sub-Eddington rate (several tens of percent of the
Eddington luminosity --- see Fig. 12). The three objects with a
slightly lower temperature, $kT_{\rm in}\sim 0.5$ keV (Fig. 12), have
nominally higher mass (20--30$M_\odot$) black holes and
$\eta$ {\simlt} 10\%, although the errors in $kT_{\rm in}$, and
consequently in $M$ and $\eta$, are relatively large.

Spectral information for less luminous objects was obtained from
analysis of the band ratio C(2--8 keV)/C(0.5--2 keV). Figs 5a and 5b
indicate that most objects have a band ratio indicative of
$1<\Gamma<2$, which is consistent with the spectra of Galactic
low-mass and high-mass X-ray binary systems. A photon index of $\Gamma
\sim 1$ is slightly harder than typically observed in Galactic black
hole candidates in their low/hard state. Sources with such hard
spectra (Fig. 5) may have larger absorbing columns than the Galactic
column or have intrinsically flat spectra.

\subsubsection{Unified Picture of the Luminous X-ray Sources in M51}

The above discussed information about spectra and variability of
individual sources does not provide strong constraints on the mass and
mass accretion rates of the compact objects. In this section, we
discuss possible unified pictures of the luminous X-ray sources in
M51 by examining their overall ensemble properties.

A recent systematic study with {\chandra} and {\xmm} of about 50
luminous X-ray sources ({\simgt}$5\times10^{38}$ {\eps}) in various
galaxies has shown that spectra of 28\% and 42\% of the sample are
fitted better by a MCD and a power law model, respectively. The rest
are well described by either a MCD or power law (Sugiho 2003). Since
most of our spectra have less good photon statistics than this sample,
most of the objects in M51 which are well fitted with either a MCD or
power law model would thus be classified into a MCD- or a power law-
type, if better signal to noise spectra were obtained. Therefore, it
is plausible that the current sample contains both classes (i.e., MCD
and power law spectra).

We now discuss three possible scenarios for the luminous sources.
Physically, power-law spectra may result from non-thermal processes,
possibly with relativistic beaming, or through Comptonization of
photons from the accretion disk.

\noindent
(a) According to the MCD interpretation (discussed in section 4.2.5),
the spectral parameters of most sources are compatible with stellar
mass black holes ({\simlt}$10M_{\odot}$) with bolometric luminosities
near or above the Eddington luminosity. This interpretation is viable
for at least some of the luminous sources in the present study based
on the statistical results above. The state with a MCD spectrum can be
identified with the high/soft state of a black hole with a thin
accretion disk (Fig. 12) or a slim disk (Fig. 13), both of which show
a MCD-like spectral shape, as inferred from observations of Galactic
black hole candidates. The objects with a power law spectrum might be
in a state with high accretion rates and significant Comptonization of
photons from the accretion disk. This state might be a very high state
which appears at a mass accretion rate between the values for the
high/soft state of thin accretion disks and the slim disk. If this is
the case, the three states can be described by a sequence of
increasing mass accretion rates: the high/soft state, an apparently
power-law state via strong Comptonization, and the slim disk.

\noindent
(b) On the other hand, the observed power law spectra may come from
black holes in a low/hard state resulting from Comptonization of disk
photons in a corona or from an advection-dominated accretion flow. In
this case, the mass accretion rate should be very low (less than a few
\% of $L_{\rm Edd}$), and then the black hole mass would be very
high. For example, if a source with $L_{\rm bol} = 5\times10^{38}$
{\eps} is radiating at $<$3\% of the Eddington luminosity, the black
hole mass becomes $>$ 100 $M_{\odot}$, in the range of IMBHs. If this
is the case, a large number of IMBHs would be present in one galaxy
since a significant fraction of the X-ray sources in M51 may be the
power-law type, based on the spectral fits and the statistical
argument of Sugiho (2003).

\noindent
(c) If relativistic beaming is significant in the power-law sources,
our line of sight must be within the solid angle of the beam $\Delta
\Omega$ ($<<4\pi$).  Then there must be a large number of the same
class which are misaligned with respect to the line of site. If the
relativistic flow is bipolar, the number of such a class is $2\pi
N_{\rm beamed}/ \Delta \Omega >> N_{\rm beamed}$, where $N_{\rm
beamed}$ is the number of sources with the beam aligned with the line
of sight. The misaligned examples might be observed as low
luminosity objects with a MCD spectrum.

Although the present data cannot exclude any of these possibilities,
the first one (a), in which a power law spectrum is a result of
Comptonization at a high accretion rate, has the advantage that most
of the objects can be explained by a sequence of fundamental
parameters (black hole mass, mass accretion rate) in a relatively
small range ($M$ $\approx$ 5 -- 10 $M_{\odot}$ and $\eta \approx$ 0.3
-- 6). On the other hand, the second and third possibilities need to
introduce another population, namely IMBHs (b) and micro quasars
(c), respectively.

\subsection{Super Soft ULXs}

% src 98 = NGC 5194 \#9
% src 34 = NGC 5195 \#5

Two sources (NGC 5194 \#9 = CXOM51~J132943.3+471135 and NGC 5195 \#5 =
CXOM51 J132958.4+471547) have very soft spectra and no photons were
detected above 1 keV. MCD models require $kT\sim0.1$ keV (Table 6),
which is unusually low for Galactic black hole candidates.  A
blackbody model provides similar temperatures (Table 9). Their
luminosities exceed $10^{38}$ {\eps} in the 0.5--8 keV band, although
the errors on the luminosities are large because of the limited energy
band used in the fits and strong coupling with the absorbing
column. The luminosities of NGC 5194 \#9 in the first observation and
NGC 5195 \#5 in the second observation exceed $10^{39}$ {\eps} (for
the best-fit MCD model), which is in the range of ULXs. Fig. 14 shows
the data and best-fit blackbody model together with the same blackbody
model after {\NH} has been set to 0, to demonstrate the effect of
absorption.

The very soft spectra ($kT\sim0.1$ keV) are similar to those observed
in super soft sources (SSSs) in our Galaxy and the Magellanic clouds
(e.g., Kahabka \& van den Heuvel 1997). The luminosities, however, are
much higher than those of the typical SSS ($10^{36}-10^{38}$ {\eps}).
Such a soft source with a large luminosity ($\sim10^{39}$ {\eps} or
higher) is known in M81 (Source N1, Swartz et al. 2002), the Antennae
(CXOANT~J120151.6--185231.9, Fabbiano et al. 2003), and M101 (P098,
Mukai et al. 2003).

SSSs often show absorption edges in their spectra (Asai et al. 1998;
Shimura 2000; Ebisawa et al. 2001a). Similar absorption features are
detected in luminous SSSs in M81 (Swartz et al. 2002). Such
features, however, are not seen in our spectra in view of the limited
photon statistics.  Higher quality spectra are needed to search for
these absorption edges.

An optically thin thermal plasma model (MEKAL) was also examined and
the results are also shown in Table 9. Although acceptable fits were
obtained for NGC 5194 \#9, the abundance is unphysically low (a few \%
of solar abundance or less) and the thermal plasma model is thus
implausible.

Low temperature ($kT\sim 0.1$ keV) and high luminosity ($L_{\rm
bol}>10^{39}$ {\eps}) are just what would be expected for accretion
disk emission around an IMBH. The MCD parameters derived from our
model fits for NGC 5194 \#9 (first observation), NGC 5194 \#9 (second
observation), and NGC 5195 \#5 (second observation) correspond to
$R_{\rm in}$ = $1.3\times 10^5$, $1.6\times 10^4$, $6.1\times 10^4$
km; $M_{\rm BH}$ = $1.5\times 10^4$, $1.8\times 10^3$, $8.5\times
10^3$ $M_{\odot}$; $\eta$=0.035, 0.018, 0.041, respectively. This MCD
interpretation has, however, two problems.  First, the inner disk
radius for NGC 5194 \#9 decreases significantly from the first to the
second observation. In a high/soft state of Galactic black hole
candidates, $R_{\rm in}$ derived by adopting the MCD model remains
constant (e.g., Ebisawa et al. 1993). The variation of $R_{\rm in}$ in
NGC 5194 \#9 suggests that a MCD interpretation is not appropriate for
this object and that the parameters derived above may be unreliable.
Similarly, a MCD model results in an implausibly large change in the
emitting region for luminous SSSs in the Antennae (Fabbiano et
al. 2003) and M101 (Mukai et al. 2003), arguing against an accretion
disk around an IMBH.
Second, these three spectra show no hard component. In the high/soft
state of Galactic black hole candidates, a hard power-law component is
always present. This component is significant at energies above 6--7
keV for a BH with $kT_{\rm in}\sim 1$ keV, for example. Such a hard
component is also seen in IMBH candidates in NGC 1313 (Miller et
al. 2003).

Mukai et al. (2003) have proposed an optically thick outflow produced by
a BH radiating near the Eddington limit for an ultraluminous SSS found in
M101 (source P098). Since the spectral properties and luminosities of 
the SSSs in M51 are quite similar to P098 in M101, the same model may be
applicable to the sources in M51.

\subsection{The Nucleus of NGC 5195}

The nucleus of NGC 5195 is classified as a LINER 2 (Ho et
al. 1997). An X-ray source (NGC 5195 \#7 = CXOM51 J132959.5+471559)
coincides with the radio core of NGC 5195 to within the positional
uncertainty ($\sim2^{\prime\prime}$) for near the edge of the CCD
chip. The X-ray position is 1\farcs5 away from the radio core
position, which is $\alpha = 13^{\rm h}29^{\rm m}59\fs5$, $\delta =
47^{\circ}15^{\prime}57^{\prime \prime}$ (J2000) at 20 cm (Ho \&
Ulvestad 2001). On the other hand, the position of the optical peak
($13^{\rm h}29^{\rm m}58\fs80$, $47^{\circ}16^{\prime}00\farcs0$
[J2000] from the Updated Zwicky Catalog [Falco et al. 1999] obtained
via the NASA Extra Galactic Database) is 7\farcs4 away from the X-ray
source.

  Ho et al. (2001) and Georgantopoulus et al. (2002) analyzed a short
(1.1 ksec exposure) {\chandra} ACIS-S3 observation of NGC 5195.  Ho et
al. (2001) measured an upper limit of $<$ 3 counts, $L_{\rm
X}<8.4\times10^{37}$ {\eps} (for a distance of 8.4 Mpc), while
Georgantopoulus et al. (2002) reported detection of an X-ray nucleus
with a luminosity of $2.4\times10^{39}$ {\eps} in the 2--10 keV
band. The X-ray nucleus found by Georgantopoulus et al. (2002)
coincides with the radio core and is most probably the same source as
we detect. The discrepancy between Ho et al.'s and Georgantopoulus et
al.'s results may be due to the different position assumed for the
nucleus in their analyses.

No long-term variability is seen between our two observations in the
full, soft, or hard energy band. No short-term variability was seen
within either of our observations. The number of counts found by
Georgantopoulos et al. (2002) (10 counts in the 0.5--7 keV band in a
1.1 ksec exposure observation in 2000 Jan.) is consistent with the
count rate we obtained in our longer observations.

The X-ray spectra obtained in 2000 and 2001 can be described by either
a power law or a MCD model (see Table 7), and their spectral
parameters are consistent with a low-luminosity AGN (LLAGN), an X-ray
binary, or a ULX. The X-ray luminosity corrected for absorption is
$\approx 8\times10^{38}$ {\eps} in the 0.5--8 keV band for a power law
model.

The ratio $\log L_{\rm X}/L_{\rm H\alpha} = 0.19$ is smaller than is
typical of AGNs (here $L_{\rm X}$ refers to the 2--10 keV band and the
H$\alpha$ luminosity is taken from Ho, Filippenko, \& Sargent 2003).
This small ratio indicates that extrapolation of the X-ray flux into
the UV assuming $f_\nu \propto \nu^{-1}$ (e.g., Ho 1999) provides
insufficient ionizing photons to account for the H$\alpha$
luminosity. Thus if the LINER 2 nucleus is powered by an AGN, it must
be heavily obscured in the 2--10 keV band (Terashima, Ho, \& Ptak
2000a; Terashima et al. 2000b).

Optical LINER spectra can be modelled in terms of photoionization by
very hot stars, particularly for objects with small
[\ion{O}{1}]$\lambda 6300$/H$\alpha$ ratios like NGC 5195 (Filippenko
\& Terlevich 1992; Shields 1992). A problem with this scenario for NGC
5195 is the absence of O stars. The optical spectrum of the nucleus of
NGC 5195 shows strong Balmer absorption lines, which indicate the
presence of numerous A-type stars and thus a ``post-starburst'' nature
(Filippenko \& Sargent 1985; Ho, Filippenko, \& Sargent 1995; Boulade
et al. 1996; Kohno et al. 2002). The age of the starburst is estimated
at about $10^9$ years. Thus, the LINER-type emission line spectrum is
not likely to result from photoionization by early type main sequence
or Wolf-Rayet (Filippenko 1996 and references therein; Barth \&
Shields 2000) stars in view of their short lifetimes.

\section{Summary}

Our findings are summarized as follows.

\begin{enumerate}

\item We have analyzed two {\chandra} observations of M51 performed in
2000 June and 2001 June. 113 X-ray sources were detected on the
ACIS-S3 CCD chip. Of these, 84 and 12 sources project
within the optical extents of NGC 5194 and NGC 5195, respectively.

\item The cumulative luminosity function of the X-ray sources in NGC
5194 with $L$(0.5 -- 8 keV)$> 10^{38}$ {\eps} has a power law form
$N(>L$(0.5 -- 8 keV)) $\propto$ $L$(0.5 -- 8 keV)$^{-\alpha}$,
with $\alpha$= 0.91, which is
inbetween the typical slopes in spiral galaxies and starburst galaxies.
The number of ultraluminous X-ray sources (ULXs), whose luminosities
are in excess of $10^{39}$ {\eps} in the 0.5--8 keV band, is seven and
two in NGC 5194 and NGC 5195, respectively. The number of ULXs in NGC
5194 is unusually high for spiral galaxies with mild star formation
activity. The most luminous object has a luminosity of
$4\times10^{39}$ {\eps} in the 0.5--8 keV band.

\item The majority of the X-ray sources are located in or near the
spiral arms. All seven of the ULXs in NGC 5194 are in, or close to, a
spiral arm, confirming an association with star forming activity. Five
out of these seven are not coincident with bright \ion{H}{2} regions,
in contrast to some ULXs observed in starburst galaxies.

\item Short term variability within each observation was detected in
three sources. About one-third of the objects in NGC 5194 or NGC 5195
varied between the two observations. Four objects exhibit large
amplitude (a factor of ten or more) variability, suggesting a
transient nature. One source (NGC 5194 \#69) exhibits a possible
period of $\approx$ 2.1 hrs in the first observation, but only two
periods are covered by the observation.

\item The spectral band ratio (counts in (2--8 keV) band /
counts in (0.5--2 keV) band) of most sources is consistent with a power
law spectrum with a photon index of 1--2. 

\item Spectral modelling was performed for 38 spectra of 26 objects,
each spectrum having more than 70 counts in an observation.  The
spectra of the two ULXs NGC 5194 \# 37 and \# 82 are better described
by a power law model than a multicolor disk blackbody (MCD) model.  In
accord with observations of ULXs in other galaxies, ULXs with a power
law spectrum constitute a class of ULXs in addition to a population
whose spectra are better described by a MCD. The power law spectra can
be interpreted in terms of (1) the low/hard state of an accreting
black hole, (2) relativistically beamed jets, or (3) emission from an
accretion disk significantly Comptonized in the very high state. No
strong constraints on the masses of the central objects and accretion
rates have been obtained from the present data.

One transient ULX (NGC 5194 \# 69) showed a drastic steepening of its
spectrum, reminiscent of Galactic soft X-ray transients.  The
intrinsic luminosity of this source in the 2--10 keV band decreased by
a factor of 3460 from the first to the second observation. Such a
drastic spectral state change and possible periodicity may provide a
link to the elaborate phenomenology of stellar mass black hole
candidates.

One ULX (NGC 5194 \# 26) displayed strong emission lines from highly
ionized species in the first observation; these lines disappeared in
the second observation, when the luminosity was 50\% higher than in
the first observation.

The spectra of the other ULXs and less luminous sources may be
described by either a power law or a MCD model. If we adopt the MCD
model, their spectral parameters are consistent with an accretion disk
around a black hole with a mass of 5--10 $M_{\odot}$ accreting at near
or above the Eddington rate.

There are three sources with very soft spectra ($kT\approx0.15$
keV). Their inferred X-ray luminosities
($4.0\times10^{38}-2.6\times10^{39}$ {\eps}) are unusually high
compared to super soft sources known so far. We discuss the idea that
this soft emission originates in an accretion disk around an
intermediate mass black hole.

\item From the spectral properties of the luminous sources (except the
sources with very soft spectra), we speculate that they are
stellar-mass black holes radiating around or above the Eddington
luminosity. In this scenario, sources with a power law spectrum can be
interpreted in terms of Comptonized accretion disk emission when
accretion rates are high. X-ray emission from sources best described
by a MCD model originates in the optically-thick geometrically-thin
accretion disk or slim disk. Alternatively, sources with power law
spectra may be in a ``low/hard'' state, in which case the black hole
mass becomes $>$ 100 $M_{\odot}$, and in the range of
intermediate-mass black holes. Another possibility is that the
emission from the power-law sources originates in a relativistic jet.

\item The LINER 2 nucleus of NGC 5195 may be powered by a heavily
obscured AGN and/or shocks.  The gas radiating the narrow emission
lines is unlikely to be photoionized by O stars.

\end{enumerate}

\acknowledgments

The authors are grateful to D.~A. Thilker for providing optical images
and a list of \ion{H}{2} regions in computer readable format. We wish
to thank K. Watarai for his slim disk calculations, and A. Kubota for
her useful input. M.~C. Miller gave valuable comments on the
manuscript.  Y.~T. is supported by the Japan Society for the Promotion
of Science. This research was supported by NASA through grants
NAG81027 and NAG81755 to the University of Maryland.

\newpage

\begin{center}

\begin{deluxetable}{ccccccccc}
\tablewidth{16cm}
	\tablecaption{Number of Detected Sources}
%\tabletypesize{\scriptsize}
\tablehead{
\colhead{Location}    & 
\multicolumn{8}{c}{Energy Band} \\
\cline{2-9}
\colhead{}		& 
\multicolumn{2}{c}{(0.5--8 keV)}	& 
\multicolumn{2}{c}{(0.5--2 keV)}	& 
\multicolumn{2}{c}{(2--8 keV)}	& 
\multicolumn{2}{c}{(Any One of the Three Bands)} \\
\colhead{}	      & 
\colhead{Detected}    & 
\colhead{Variable}    & 
\colhead{Detected}    & 
\colhead{Variable}    & 
\colhead{Detected}    & 
\colhead{Variable}    & 
\colhead{Detected}    & 
\colhead{Variable}    \\
}
\startdata
S3 chip		& 113 & 32 & 102 & 30 & 48 & 15 & 113 & 36\\	
NGC 5194	& 84 & 26 & 76 & 22 & 34 & 14 & 84 & 28\\
NGC 5195	& 12 & 3  & 11 & 3  & 8  & 0  & 12 & 3\\ 
\enddata
\end{deluxetable}

\end{center}

% radec5194rv4.tex 

\setcounter{table}{1}

\begin{deluxetable}{ccccrrrrrrrrr}
\rotate
\tablewidth{22.5cm}
%\tabletypesize{\scriptsize}
\tablecaption{Sources spatially coincident with NGC 5194}
\tablehead{ 
\colhead{No.}    & 
\colhead{RA}    & 
\colhead{Dec.}  & 
\colhead{CXO Name}      &  
\multicolumn{3}{c}{Counts (Observations 1+2)}  & 
\multicolumn{3}{c}{Counts (Observation 1)}  & 
\multicolumn{3}{c}{Counts (Observation 2)}  \\ 
\colhead{}    & 
\colhead{(J2000)}    & 
\colhead{(J2000)}    &
\colhead{CXOM51}    & 
\colhead{0.5--8 keV}    &  
\colhead{0.5--2 keV}    &  
\colhead{2--8 keV}    &  
\colhead{0.5--8 keV}    &  
\colhead{0.5--2 keV}    &  
\colhead{2--8 keV}    &  
\colhead{0.5--8 keV}    &  
\colhead{0.5--2 keV}    &  
\colhead{2--8 keV}    \\
\colhead{(1)}   &
\colhead{(2)}   &
\colhead{(3)}   &
\colhead{(4)}   &
\colhead{(5)}   &
\colhead{(6)}   &
\colhead{(7)}   &
\colhead{(8)}   &
\colhead{(9)}   &
\colhead{(10)}  &
\colhead{(11)}  &
\colhead{(12)}  &
\colhead{(13)}  }
\startdata
     1	  &  13 29 35.69  &  47 12 01.1  &   J132935.7+471201  &  15.2$\pm$5.4  &  $<$9.4  &  11.6$\pm$4.7  &  $<$14.0  &  $<$4.3  &  6.6$\pm$3.8  &  8.2$\pm$4.3  &  $<$8.7  &  $<$11.2  \\
     2	  &  13 29 36.55  &  47 11 05.4  &   J132936.5+471105  &  8.8$\pm$4.6  &  6.7$\pm$2.6  &  $<$7.9  &  $<$6.6  &  $<$5.7  &  $<$4.4  &  $<$14.3  &  $<$11.1  &  $<$7.0  \\
     3	  &  13 29 38.92  &  47 13 23.9  &   J132938.9+471324  &  54.6$\pm$8.8  &  37.5$\pm$7.3  &  17.4$\pm$4.2  &  16.6$\pm$4.1  &  12.9$\pm$3.6  &  $<$8.9  &  38.5$\pm$7.6  &  24.9$\pm$6.2  &  13.6$\pm$3.7  \\
     4	  &  13 29 38.98  &  47 11 03.6  &   J132939.0+471104  &  26.3$\pm$5.2  &  16.6$\pm$4.1  &  9.7$\pm$3.2  &  14.7$\pm$5.0  &  8.9$\pm$3.0  &  $<$12.2  &  12.7$\pm$3.6  &  7.9$\pm$2.8  &  $<$10.4  \\
     5	  &  13 29 39.45  &  47 12 43.7  &   J132939.5+471244  &  427$\pm$21  &  308$\pm$18  &  120$\pm$11  &  244$\pm$16  &  170$\pm$13  &  73.6$\pm$8.6  &  184$\pm$14  &  138$\pm$12  &  46.5$\pm$6.9  \\
     6	  &  13 29 39.96  &  47 12 36.9  &   J132940.0+471237  &  276$\pm$17  &  266$\pm$16  &  10.4$\pm$4.4  &  226$\pm$15  &  217$\pm$15  &  8.8$\pm$4.1  &  50.0$\pm$7.1  &  48.3$\pm$7.0  &  $<$5.9  \\
     7	  &  13 29 41.65  &  47 10 52.3  &   J132941.6+471052  &  20.4$\pm$4.6  &  16.6$\pm$4.1  &  $<$8.9  &  $<$9.0  &  $<$6.1  &  $<$6.2  &  16.6$\pm$4.1  &  14.7$\pm$3.9  &  $<$5.9  \\
     8	  &  13 29 42.51  &  47 10 42.7  &   J132942.5+471043  &  16.7$\pm$5.3  &  13.1$\pm$4.8  &  $<$9.1  &  $<$10.3  &  $<$9.0  &  $<$4.6  &  12.3$\pm$4.7  &  9.5$\pm$4.3  &  $<$7.7  \\
     9	  &  13 29 43.30  &  47 11 34.7  &   J132943.3+471135  &  365$\pm$19  &  364$\pm$19  &  $<$5.8  &  147$\pm$12  &  146$\pm$12  &  $<$4.6  &  219$\pm$15  &  218$\pm$15  &  $<$4.5  \\
    10	  &  13 29 44.05  &  47 11 56.4  &   J132944.1+471156  &  18.8$\pm$5.7  &  17.3$\pm$5.4  &  $<$5.8  &  8.2$\pm$4.1  &  7.3$\pm$4.0  &  $<$4.6  &  10.6$\pm$4.6  &  9.9$\pm$4.4  &  $<$4.5  \\
    11	  &  13 29 44.19  &  47 10 20.3  &   J132944.2+471020  &  8.1$\pm$4.4  &  8.7$\pm$4.4  &  $<$4.3  &  $<$8.3  &  $<$8.6  &  $<$4.6  &  $<$11.7  &  $<$12.0  &  $<$4.5  \\
    12	  &  13 29 44.55  &  47 13 56.7  &   J132944.5+471357  &  72.8$\pm$8.6  &  55.4$\pm$8.6  &  18.6$\pm$4.4  &  14.8$\pm$3.9  &  10.5$\pm$4.4  &  $<$10.0  &  58.1$\pm$7.7  &  43.9$\pm$7.8  &  13.7$\pm$3.7  \\
    13	  &  13 29 44.87  &  47 11 28.6  &   J132944.9+471129  &  11.8$\pm$4.8  &  7.3$\pm$4.1  &  $<$10.3  &  $<$10.0  &  $<$8.8  &  $<$4.6  &  $<$14.9  &  $<$9.8  &  $<$9.1  \\
    14	  &  13 29 45.51  &  47 11 51.0  &   J132945.5+471151  &  30.0$\pm$5.6  &  14.3$\pm$3.9  &  15.7$\pm$4.0  &  12.7$\pm$3.6  &  6.8$\pm$2.6  &  5.9$\pm$2.4  &  17.3$\pm$4.2  &  7.6$\pm$2.8  &  9.8$\pm$3.2  \\
    15	  &  13 29 45.52  &  47 11 25.2  &   J132945.5+471125  &  11.3$\pm$3.5  &  8.5$\pm$3.0  &  $<$7.4  &  $<$4.2  &  $<$4.3  &  $<$4.6  &  10.5$\pm$3.3  &  8.7$\pm$3.0  &  $<$6.0  \\
    16	  &  13 29 45.94  &  47 10 55.7  &   J132945.9+471056  &  9.0$\pm$3.2  &  7.5$\pm$2.8  &  $<$5.8  &  $<$7.0  &  $<$7.2  &  $<$4.6  &  $<$13.6  &  $<$11.2  &  $<$6.0  \\
    17	  &  13 29 46.11  &  47 10 42.3  &   J132946.1+471042  &  28.6$\pm$5.5  &  25.0$\pm$5.1  &  $<$8.9  &  26.6$\pm$5.2  &  23.7$\pm$4.9  &  $<$7.8  &  $<$6.4  &  $<$5.4  &  $<$4.5  \\
    18	  &  13 29 47.02  &  47 11 04.3  &   J132947.0+471104  &  11.8$\pm$4.8  &  12.3$\pm$4.8  &  $<$4.4  &  $<$8.6  &  $<$8.8  &  $<$4.6  &  8.6$\pm$4.3  &  8.9$\pm$4.3  &  $<$4.5  \\
    19	  &  13 29 47.51  &  47 13 01.3  &   J132947.5+471301  &  17.8$\pm$4.4  &  10.5$\pm$3.3  &  7.5$\pm$4.0  &  16.2$\pm$5.2  &  9.3$\pm$4.3  &  6.9$\pm$2.6  &  $<$5.1  &  $<$4.1  &  $<$4.5  \\
    20	  &  13 29 48.79  &  47 11 21.4  &   J132948.8+471121  &  67.8$\pm$9.4  &  50.3$\pm$8.3  &  17.7$\pm$4.2  &  18.2$\pm$5.4  &  15.3$\pm$5.1  &  $<$7.8  &  49.6$\pm$8.2  &  34.9$\pm$7.1  &  14.8$\pm$3.9  \\
    21	  &  13 29 49.03  &  47 10 53.3  &   J132949.0+471053  &  143$\pm$12  &  105$\pm$10  &  36.5$\pm$6.1  &  $<$5.6  &  $<$5.7  &  $<$4.6  &  142$\pm$12  &  104$\pm$10  &  36.7$\pm$6.1  \\
    22	  &  13 29 49.15  &  47 12 57.0  &   J132949.2+471257  &  14.0$\pm$3.9  &  $<$3.9  &  13.7$\pm$3.7  &  8.7$\pm$3.0  &  $<$4.3  &  8.9$\pm$3.0  &  $<$10.8  &  $<$4.1  &  $<$10.5  \\
    23	  &  13 29 50.06  &  47 14 19.9  &   J132950.1+471420  &  17.7$\pm$5.7  &  16.0$\pm$5.3  &  $<$6.5  &  $<$12.5  &  $<$12.9  &  $<$4.4  &  $<$19.4  &  8.8$\pm$3.0  &  $<$7.0  \\
    24	  &  13 29 50.08  &  47 11 39.3  &   J132950.1+471139  &  32.3$\pm$6.2  &  31.8$\pm$6.1  &  $<$4.2  &  12.0$\pm$3.9  &  12.5$\pm$3.9  &  $<$4.5  &  21.8$\pm$5.0  &  20.7$\pm$4.9  &  $<$4.3  \\
    25	  &  13 29 50.35  &  47 13 22.7  &   J132950.4+471323  &  22.8$\pm$6.1  &  22.3$\pm$6.0  &  $<$4.4  &  $<$12.8  &  $<$11.6  &  $<$4.6  &  16.6$\pm$5.3  &  16.9$\pm$5.3  &  $<$4.5  \\
    26	  &  13 29 50.67  &  47 11 55.1  &   J132950.7+471155  &  377$\pm$20  &  52.2$\pm$9.0  &  325$\pm$18  &  104$\pm$11  &  $<$19.2  &  95.6$\pm$9.8  &  273$\pm$17  &  42.8$\pm$8.1  &  229$\pm$15  \\
    27	  &  13 29 50.82  &  47 10 31.3  &   J132950.8+471031  &  135$\pm$13  &  117$\pm$11  &  17.5$\pm$5.3  &  71.2$\pm$9.5  &  61.4$\pm$7.9  &  9.8$\pm$4.3  &  63.6$\pm$9.1  &  55.9$\pm$8.6  &  7.7$\pm$4.0  \\
    28	  &  13 29 51.36  &  47 10 32.4  &   J132951.4+471032  &  238$\pm$16  &  155$\pm$13  &   83$\pm$10  &  73.2$\pm$9.6  &  48.4$\pm$8.1  &  24.8$\pm$6.1  &  165$\pm$13  &  107$\pm$10  &  58.6$\pm$7.7  \\
    29	  &  13 29 51.78  &  47 11 52.0  &   J132951.8+471152  &  36.3$\pm$8.1  &  36.2$\pm$8.0  &  $<$4.2  &  13.1$\pm$5.3  &  13.4$\pm$5.3  &  $<$4.5  &  24.2$\pm$6.7  &  23.8$\pm$6.6  &  $<$4.3  \\
    30	  &  13 29 52.07  &  47 11 26.8  &   J132952.1+471127  &  24.3$\pm$7.1  &  25.2$\pm$7.1  &  $<$4.2  &  $<$16.7  &  $<$17.6  &  $<$4.5  &  18.8$\pm$4.9  &  19.1$\pm$4.9  &  $<$4.3  \\
    31	  &  13 29 52.12  &  47 12 12.8  &   J132952.1+471213  &  22.1$\pm$5.3  &  21.7$\pm$5.2  &  $<$5.7  &  8.2$\pm$4.6  &  $<$15.2  &  $<$4.7  &  12.6$\pm$3.9  &  12.8$\pm$3.9  &  $<$4.4  \\
    32	  &  13 29 52.23  &  47 11 29.6  &   J132952.2+471130  &  25.3$\pm$7.2  &  24.2$\pm$7.1  &  $<$5.5  &  $<$16.7  &  $<$17.6  &  $<$4.5  &  17.2$\pm$6.1  &  15.8$\pm$5.9  &  $<$5.8  \\
    33	  &  13 29 52.73  &  47 10 51.2  &   J132952.7+471051  &  34.0$\pm$6.0  &  15.8$\pm$4.1  &  18.5$\pm$5.4  &  14.2$\pm$5.0  &  $<$10.2  &  9.9$\pm$3.2  &  19.9$\pm$4.6  &  10.9$\pm$4.6  &  8.7$\pm$3.0  \\
    34	  &  13 29 52.73  &  47 11 21.4  &   J132952.7+471121  &  16.1$\pm$4.6  &  15.0$\pm$4.5  &  $<$4.2  &  $<$6.4  &  $<$5.6  &  $<$4.5  &  12.7$\pm$4.0  &  12.8$\pm$4.0  &  $<$4.3  \\
    35	  &  13 29 52.75  &  47 10 42.6  &   J132952.7+471043  &  11.8$\pm$4.8  &  $<$15.9  &  $<$8.9  &  8.2$\pm$4.1  &  $<$13.0  &  $<$6.2  &  $<$9.3  &  $<$6.7  &  $<$6.0  \\
    36	  &  13 29 52.79  &  47 12 44.9  &   J132952.8+471245  &  42.8$\pm$7.8  &  28.3$\pm$6.5  &  14.5$\pm$5.0  &  30.2$\pm$6.6  &  20.4$\pm$5.7  &  9.9$\pm$3.2  &  12.6$\pm$4.8  &  $<$15.2  &  $<$10.5  \\
    37	  &  13 29 53.31  &  47 10 42.3  &   J132953.3+471042  &  581$\pm$24  &  429$\pm$21  &  151$\pm$12  &  $<$5.6  &  $<$5.7  &  $<$4.6  &  581$\pm$24  &  429$\pm$21  &  151$\pm$12  \\
    38	  &  13 29 53.53  &  47 11 32.9  &   J132953.5+471133  &  27.3$\pm$7.4  &  19.2$\pm$6.6  &  8.1$\pm$4.1  &  $<$21.0  &  $<$14.4  &  $<$10.5  &  15.5$\pm$4.6  &  11.9$\pm$4.1  &  $<$8.9  \\
    39	  &  13 29 53.54  &  47 11 26.4  &   J132953.5+471126  &  36.2$\pm$6.6  &  32.1$\pm$6.2  &  $<$11.4  &  12.6$\pm$3.9  &  11.8$\pm$3.7  &  $<$6.0  &  22.7$\pm$5.2  &  19.8$\pm$4.8  &  $<$8.9  \\
    40	  &  13 29 53.62  &  47 13 22.8  &   J132953.6+471323  &  12.0$\pm$3.6  &  12.3$\pm$3.6  &  $<$4.4  &  $<$14.2  &  $<$14.3  &  $<$4.6  &  $<$10.8  &  $<$11.2  &  $<$4.5  \\
    41	  &  13 29 53.72  &  47 14 35.9  &   J132953.7+471436  &  629$\pm$26  &  445$\pm$22  &  184$\pm$15  &  227$\pm$16  &  165$\pm$14  &  62.1$\pm$9.0  &  401$\pm$21  &  279$\pm$18  &  122$\pm$12  \\
    42	  &  13 29 53.79  &  47 14 31.7  &   J132953.8+471432  &  276$\pm$18  &  217$\pm$16  &  58.9$\pm$8.9  &  114$\pm$12  &   92$\pm$11  &  $<$33.8  &  164$\pm$14  &  127$\pm$12  &  36.8$\pm$7.2  \\
    43	  &  13 29 53.93  &  47 10 31.3  &   J132953.9+471031  &  12.7$\pm$3.7  &  13.3$\pm$5.0  &  $<$4.4  &  $<$4.2  &  $<$4.3  &  $<$4.6  &  12.6$\pm$4.8  &  12.9$\pm$4.8  &  $<$4.5  \\
    44	  &  13 29 54.00  &  47 09 23.3  &   J132954.0+470923  &  9.6$\pm$4.4  &  8.2$\pm$4.1  &  $<$5.8  &  $<$7.4  &  $<$6.1  &  $<$4.6  &  $<$14.1  &  $<$13.1  &  $<$4.5  \\
    45	  &  13 29 54.16  &  47 11 30.2  &   J132954.2+471130  &  18.3$\pm$6.6  &  16.2$\pm$6.4  &  $<$6.9  &  $<$16.7  &  $<$14.4  &  $<$6.0  &  11.2$\pm$5.4  &  9.8$\pm$5.2  &  $<$5.8  \\
    46	  &  13 29 54.18  &  47 11 37.0  &   J132954.2+471137  &  111$\pm$12  &  68.0$\pm$8.9  &  43.1$\pm$7.7  &  44.1$\pm$6.9  &  26.7$\pm$5.5  &  17.7$\pm$4.2  &  68.2$\pm$9.8  &  42.8$\pm$8.1  &  25.5$\pm$6.2  \\
    47	  &  13 29 54.24  &  47 13 00.2  &   J132954.2+471300  &  47.2$\pm$7.0  &  15.0$\pm$4.0  &  32.5$\pm$6.8  &  18.5$\pm$4.4  &  7.7$\pm$2.8  &  10.9$\pm$3.3  &  29.0$\pm$5.5  &  7.4$\pm$2.8  &  21.7$\pm$4.7  \\
    48	  &  13 29 54.37  &  47 11 21.5  &   J132954.4+471122  &  26.8$\pm$5.7  &  27.5$\pm$5.7  &  $<$4.2  &  11.8$\pm$3.6  &  11.9$\pm$3.6  &  $<$4.5  &  13.9$\pm$4.1  &  14.2$\pm$4.1  &  $<$4.3  \\
    49	  &  13 29 54.53  &  47 09 21.6  &   J132954.5+470922  &  69.2$\pm$8.4  &  45.7$\pm$6.9  &  23.6$\pm$4.9  &  25.5$\pm$5.1  &  17.7$\pm$4.2  &  7.9$\pm$2.8  &  42.0$\pm$6.6  &  26.4$\pm$5.2  &  15.8$\pm$4.0  \\
    50	  &  13 29 54.96  &  47 09 22.4  &   J132955.0+470922  &  47.6$\pm$7.0  &  38.1$\pm$6.2  &  9.5$\pm$4.3  &  16.5$\pm$5.2  &  12.7$\pm$3.6  &  $<$9.2  &  31.2$\pm$5.7  &  25.4$\pm$6.2  &  5.9$\pm$2.4  \\
    51	  &  13 29 54.99  &  47 11 02.2  &   J132955.0+471102  &  12.5$\pm$3.7  &  10.8$\pm$3.5  &  $<$5.8  &  7.7$\pm$2.8  &  $<$11.6  &  $<$6.2  &  $<$10.8  &  $<$11.2  &  $<$4.5  \\
    52	  &  13 29 55.13  &  47 10 42.2  &   J132955.1+471042  &  29.8$\pm$6.7  &  29.3$\pm$6.6  &  $<$4.4  &  8.6$\pm$3.0  &  7.3$\pm$4.0  &  $<$4.6  &  21.6$\pm$5.9  &  21.9$\pm$5.9  &  $<$4.5  \\
    53	  &  13 29 55.24  &  47 10 46.2  &   J132955.2+471046  &  8.1$\pm$3.0  &  8.3$\pm$3.0  &  $<$4.4  &  $<$7.0  &  $<$7.2  &  $<$4.6  &  $<$10.8  &  $<$11.2  &  $<$4.5  \\
    54	  &  13 29 55.45  &  47 11 43.5  &   J132955.5+471143  &  49.6$\pm$7.5  &  45.4$\pm$7.2  &  $<$10.0  &  11.3$\pm$3.6  &  11.5$\pm$3.6  &  $<$4.5  &  37.2$\pm$7.8  &  32.8$\pm$7.4  &  $<$10.3  \\
    55	  &  13 29 55.47  &  47 14 02.0  &   J132955.5+471402  &  16.6$\pm$4.2  &  16.2$\pm$5.5  &  $<$4.1  &  10.5$\pm$4.6  &  9.8$\pm$4.4  &  $<$4.5  &  $<$12.8  &  $<$13.5  &  $<$4.3  \\
    56	  &  13 29 55.63  &  47 09 18.7  &   J132955.6+470919  &  7.6$\pm$4.1  &  $<$11.4  &  $<$7.4  &  $<$5.8  &  $<$4.5  &  $<$4.6  &  $<$12.8  &  $<$10.3  &  $<$6.0  \\
    57	  &  13 29 55.66  &  47 09 10.6  &   J132955.7+470911  &  12.8$\pm$5.2  &  12.4$\pm$5.0  &  $<$5.0  &  $<$4.0  &  $<$4.2  &  $<$4.3  &  13.3$\pm$5.1  &  12.3$\pm$4.8  &  $<$5.4  \\
    58	  &  13 29 55.70  &  47 10 43.6  &   J132955.7+471044  &  14.8$\pm$5.2  &  14.3$\pm$5.1  &  $<$4.4  &  $<$12.8  &  $<$13.0  &  $<$4.6  &  9.1$\pm$3.2  &  8.4$\pm$3.0  &  $<$4.5  \\
    59	  &  13 29 55.86  &  47 11 44.5  &   J132955.9+471145  &  42.5$\pm$6.9  &  41.5$\pm$6.9  &  $<$4.2  &  13.8$\pm$3.9  &  13.0$\pm$3.7  &  $<$4.5  &  27.0$\pm$5.6  &  27.4$\pm$5.6  &  $<$4.3  \\
    60	  &  13 29 56.06  &  47 13 50.9  &   J132956.1+471351  &  14.8$\pm$4.0  &  15.1$\pm$4.0  &  $<$4.2  &  $<$12.5  &  $<$11.3  &  $<$4.6  &  8.5$\pm$4.3  &  9.0$\pm$4.3  &  $<$4.4  \\
    61	  &  13 29 56.15  &  47 12 36.6  &   J132956.1+471237  &  18.8$\pm$5.7  &  18.3$\pm$5.6  &  $<$4.4  &  $<$10.0  &  $<$8.8  &  $<$4.6  &  14.6$\pm$5.1  &  14.9$\pm$5.1  &  $<$4.5  \\
    62	  &  13 29 57.47  &  47 10 37.1  &   J132957.5+471037  &  18.8$\pm$5.7  &  18.9$\pm$4.5  &  $<$4.4  &  $<$12.8  &  $<$13.0  &  $<$4.6  &  13.2$\pm$3.7  &  12.4$\pm$3.6  &  $<$4.5  \\
    63	  &  13 29 57.57  &  47 10 48.3  &   J132957.6+471048  &  400$\pm$20  &  318$\pm$18  &   81$\pm$10  &  118$\pm$11  &   92$\pm$11  &  25.8$\pm$6.2  &  282$\pm$17  &  226$\pm$15  &  55.6$\pm$7.5  \\
    64	  &  13 29 57.62  &  47 12 06.2  &   J132957.6+471206  &   95$\pm$11  &  73.9$\pm$8.7  &  20.6$\pm$4.6  &  92.1$\pm$9.6  &  72.3$\pm$8.5  &  19.8$\pm$5.6  &  $<$7.8  &  $<$6.7  &  $<$4.5  \\
    65	  &  13 29 58.36  &  47 13 20.5  &   J132958.4+471321  &  76.1$\pm$8.8  &  53.6$\pm$7.4  &  22.6$\pm$4.8  &  32.4$\pm$5.7  &  20.6$\pm$4.6  &  11.9$\pm$3.5  &  43.8$\pm$6.7  &  33.1$\pm$6.9  &  10.7$\pm$4.4  \\
    66	  &  13 29 58.72  &  47 10 29.9  &   J132958.7+471030  &  53.8$\pm$7.4  &  27.2$\pm$5.3  &  26.5$\pm$5.2  &  32.6$\pm$5.7  &  16.4$\pm$5.2  &  15.9$\pm$4.0  &  21.2$\pm$4.7  &  10.6$\pm$3.3  &  10.7$\pm$4.4  \\
    67	  &  13 29 59.02  &  47 14 34.2  &   J132959.0+471434  &  12.5$\pm$5.2  &  $<$15.0  &  $<$11.7  &  11.4$\pm$4.7  &  $<$14.0  &  $<$10.2  &  $<$6.5  &  $<$5.0  &  $<$5.2  \\
    68	  &  13 30 00.56  &  47 11 36.0  &   J133000.6+471136  &  13.2$\pm$5.1  &  $<$20.7  &  $<$5.7  &  $<$12.5  &  $<$11.4  &  $<$4.6  &  $<$14.6  &  $<$13.7  &  $<$4.5  \\
    69	  &  13 30 01.02  &  47 13 44.0  &   J133001.0+471344  &  553$\pm$24  &  374$\pm$20  &  180$\pm$13  &  507$\pm$23  &  336$\pm$18  &  171$\pm$13  &  46.3$\pm$7.0  &  38.0$\pm$6.3  &  $<$13.3  \\
    70	  &  13 30 01.12  &  47 13 32.9  &   J133001.1+471333  &  12.4$\pm$5.1  &  12.2$\pm$3.7  &  $<$4.4  &  $<$7.8  &  $<$8.0  &  $<$4.6  &  9.8$\pm$4.6  &  9.1$\pm$3.2  &  $<$4.5  \\
    71	  &  13 30 01.27  &  47 12 44.2  &   J133001.3+471244  &  15.2$\pm$4.1  &  $<$14.1  &  $<$11.8  &  $<$7.0  &  $<$5.7  &  $<$4.6  &  10.2$\pm$3.3  &  $<$12.1  &  $<$10.6  \\
    72	  &  13 30 01.88  &  47 08 43.7  &   J133001.9+470844  &  10.4$\pm$4.7  &  11.6$\pm$4.7  &  $<$4.0  &  $<$9.9  &  $<$10.4  &  $<$4.3  &  $<$13.3  &  $<$14.0  &  $<$4.3  \\
    73	  &  13 30 02.10  &  47 12 38.1  &   J133002.1+471238  &  17.6$\pm$4.4  &  15.0$\pm$4.0  &  $<$7.4  &  10.1$\pm$4.4  &  10.3$\pm$4.4  &  $<$4.6  &  $<$13.2  &  $<$9.2  &  $<$7.6  \\
    74	  &  13 30 02.13  &  47 09 00.6  &   J133002.1+470901  &  40.9$\pm$6.5  &  29.6$\pm$6.6  &  10.6$\pm$4.6  &  $<$5.5  &  $<$5.9  &  $<$4.4  &  40.2$\pm$6.4  &  28.1$\pm$6.5  &  11.0$\pm$4.6  \\
    75	  &  13 30 04.10  &  47 10 03.5  &   J133004.1+471003  &  24.2$\pm$6.5  &  24.7$\pm$6.4  &  $<$3.9  &  13.4$\pm$5.0  &  12.9$\pm$4.9  &  $<$4.4  &  10.8$\pm$4.9  &  11.9$\pm$4.9  &  $<$4.1  \\
    76	 &  13 30 04.32  &  47 13 20.8  &   J133004.3+471321  &  120$\pm$11  &  78.2$\pm$9.0  &  41.5$\pm$7.6  &  $<$9.1  &  $<$8.2  &  $<$4.4  &  115$\pm$11  &  74.3$\pm$8.7  &  40.9$\pm$7.5  \\
    77	 &  13 30 04.47  &  47 10 31.4  &   J133004.5+471031  &  31.9$\pm$5.7  &  30.3$\pm$5.6  &  $<$5.9  &  14.8$\pm$5.0  &  12.8$\pm$4.7  &  $<$6.3  &  17.4$\pm$4.2  &  17.6$\pm$4.2  &  $<$4.5  \\
    78	 &  13 30 04.66  &  47 14 17.0  &   J133004.7+471417  &  64.1$\pm$9.4  &  48.0$\pm$7.1  &  15.9$\pm$5.2  &  31.6$\pm$6.8  &  23.8$\pm$6.1  &  $<$14.8  &  33.6$\pm$5.9  &  25.9$\pm$5.2  &  8.2$\pm$4.1  \\
    79	 &  13 30 05.82  &  47 10 31.9  &   J133005.8+471032  &  13.1$\pm$3.7  &  13.4$\pm$3.7  &  $<$4.1  &  $<$15.6  &  $<$16.0  &  $<$4.4  &  $<$7.5  &  $<$8.1  &  $<$4.2  \\
    80	 &  13 30 06.04  &  47 14 04.0  &   J133006.0+471404  &  24.3$\pm$6.5  &  23.7$\pm$6.3  &  $<$5.2  &  $<$17.2  &  $<$16.4  &  $<$4.3  &  15.0$\pm$5.4  &  14.8$\pm$5.2  &  $<$4.2  \\
    81	 &  13 30 06.27  &  47 10 16.4  &   J133006.3+471016  &  9.6$\pm$4.9  &  $<$11.7  &  $<$10.9  &  $<$9.1  &  $<$5.3  &  $<$7.5  &  $<$13.4  &  $<$10.2  &  $<$6.9  \\
    82	 &  13 30 07.56  &  47 11 05.9  &   J133007.6+471106  &  1680$\pm$42  &  1367$\pm$38  &  312$\pm$18  &  831$\pm$30  &  688$\pm$26  &  139$\pm$13  &  845$\pm$29  &  673$\pm$26  &  174$\pm$14  \\
    83	 &  13 30 08.89  &  47 12 18.5  &   J133008.9+471218  &  29.0$\pm$5.6  &  24.8$\pm$5.1  &  $<$9.9  &  $<$9.2  &  $<$8.2  &  $<$4.4  &  24.0$\pm$5.0  &  20.3$\pm$4.6  &  $<$9.0  \\
    84	   &  13 30 11.03  &  47 10 40.7  &   J133011.0+471041  &  89.3$\pm$9.5  &  65.1$\pm$9.2  &  24.6$\pm$6.2  &  32.1$\pm$6.8  &  23.6$\pm$6.0  &  8.5$\pm$4.1  &  57.6$\pm$8.7  &  41.6$\pm$6.5  &  16.1$\pm$5.2  \\
\enddata
\end{deluxetable}

% radec5195rv4.tex 

\setcounter{table}{2}

\begin{deluxetable}{ccccrrrrrrrrr}
\rotate
\tablewidth{22.5cm}
%\tabletypesize{\scriptsize}
\tablecaption{Sources spatially coincident with NGC 5195}
\tablehead{ 
\colhead{No.}    & 
\colhead{RA}    & 
\colhead{Dec.}  & 
\colhead{CXO Name}      &  
\multicolumn{3}{c}{Counts (Observations 1+2)}  & 
\multicolumn{3}{c}{Counts (Observation 1)}  & 
\multicolumn{3}{c}{Counts (Observation 2)}  \\ 
\colhead{}    & 
\colhead{(J2000)}    & 
\colhead{(J2000)}    &
\colhead{CXOM51}    & 
\colhead{0.5--8 keV}    &  
\colhead{0.5--2 keV}    &  
\colhead{2--8 keV}    &  
\colhead{0.5--8 keV}    &  
\colhead{0.5--2 keV}    &  
\colhead{2--8 keV}    &  
\colhead{0.5--8 keV}    &  
\colhead{0.5--2 keV}    &  
\colhead{2--8 keV}    \\
\colhead{(1)}   &
\colhead{(2)}   &
\colhead{(3)}   &
\colhead{(4)}   &
\colhead{(5)}   &
\colhead{(6)}   &
\colhead{(7)}   &
\colhead{(8)}   &
\colhead{(9)}   &
\colhead{(10)}  &
\colhead{(11)}  &
\colhead{(12)}  &
\colhead{(13)}  }
\startdata
     1	&  13 29 52.80  &  47 16 44.5  &  J132952.8+471644  &  157$\pm$13  &  103$\pm$11  &  54.7$\pm$8.7  &  55.7$\pm$8.7  &  38.9$\pm$7.4  &  16.8$\pm$5.4 
 &  102$\pm$11  &  64.0$\pm$9.2  &  37.8$\pm$7.4  \\
     2	&  13 29 55.53  &  47 15 55.4  &  J132955.5+471555  &  22.2$\pm$5.3  &  $<$8.4  &  20.8$\pm$6.0  &  $<$12.2  &  $<$6.2  &  $<$9.8 
 &  17.0$\pm$6.1  &  $<$6.4  &  16.9$\pm$5.5  \\
     3	&  13 29 57.06  &  47 15 56.5  &  J132957.1+471556  &  25.5$\pm$8.8  &  19.6$\pm$7.9  &  $<$13.4  &  $<$18.8  &  $<$15.4  &  $<$8.7 
 &  14.0$\pm$7.3  &  12.4$\pm$6.7  &  $<$7.3  \\
     4	&  13 29 57.48  &  47 16 12.1  &  J132957.5+471612  &  256$\pm$17  &  178$\pm$16  &  74.1$\pm$9.9  &  100$\pm$12  &   69$\pm$10  &  30.5$\pm$6.7 
 &  153$\pm$13  &  109$\pm$13  &  43.5$\pm$7.8  \\
     5	&  13 29 58.38  &  47 15 47.4  &  J132958.4+471547  &  124$\pm$14  &  120$\pm$14  &  $<$9.3  &  50.6$\pm$9.2  &  49.0$\pm$9.1  &  $<$6.4 
 &   75$\pm$11  &   74$\pm$11  &  $<$6.9  \\
     6	&  13 29 58.74  &  47 16 25.5  &  J132958.7+471625  &   94$\pm$13  &   88$\pm$12  &  $<$13.6  &  $<$12.0  &  $<$12.0  &  $<$4.0 
 &   92$\pm$12  &   85$\pm$11  &  $<$15.1  \\
     7	&  13 29 59.53  &  47 15 58.5  &  J132959.5+471559  &  522$\pm$25  &  390$\pm$22  &  132$\pm$13  &  176$\pm$15  &  130$\pm$13  &  45.5$\pm$7.9 
 &  345$\pm$20  &  258$\pm$18  &   88$\pm$11  \\
     8	&  13 29 59.79  &  47 15 40.8  &  J132959.8+471541  &   89$\pm$13  &   66$\pm$11  &  22.1$\pm$6.2  &  29.6$\pm$7.9  &  25.0$\pm$7.4  &  $<$10.8 
 &   59$\pm$10  &  41.5$\pm$9.2  &  $<$28.3  \\
     9	&  13 30 00.62  &  47 15 30.5  &  J133000.6+471530  &  19.5$\pm$5.2  &  18.8$\pm$5.0  &  $<$6.7  &  22.5$\pm$5.3  &  20.3$\pm$4.9  &  $<$6.7 
 &  $<$8.0  &  $<$7.0  &  $<$4.1  \\
    10	&  13 30 02.02  &  47 15 06.3  &  J133002.0+471506  &   96$\pm$11  &  68.0$\pm$9.4  &  28.1$\pm$6.6  &  34.6$\pm$7.1  &  26.3$\pm$6.3  &  8.2$\pm$4.1 
 &  61.5$\pm$9.1  &  41.7$\pm$7.6  &  19.7$\pm$5.7  \\
    11	&  13 30 03.76  &  47 16 11.2  &  J133003.8+471611  &   85$\pm$11  &   75$\pm$10  &  9.9$\pm$4.8  &  $<$7.9  &  $<$6.2  &  $<$5.4 
 &   83$\pm$11  &  74.1$\pm$9.9  &  8.9$\pm$4.5  \\
    12	&  13 30 06.03  &  47 15 42.4  &  J133006.0+471542  &  547$\pm$25  &  386$\pm$21  &  161$\pm$14  &  200$\pm$15  &  150$\pm$13  &  50.8$\pm$8.3  &  347$\pm$20  &  237$\pm$17  &  110$\pm$12  \\
\enddata
\end{deluxetable}

% radecothersrv4.tex

\setcounter{table}{3}

\begin{deluxetable}{ccccrrrrrrrrr}
\rotate
\tablewidth{22.5cm}
%\tabletypesize{\scriptsize}
\tablecaption{Other Sources}
\tablehead{
\colhead{No.}    & 
\colhead{RA}    & 
\colhead{Dec.}  & 
\colhead{CXO Name}      &  
\multicolumn{3}{c}{Counts (Observations 1+2)}  & 
\multicolumn{3}{c}{Counts (Observation 1)}  & 
\multicolumn{3}{c}{Counts (Observation 2)}  \\ 
\colhead{}    & 
\colhead{(J2000)}    & 
\colhead{(J2000)}    &
\colhead{CXOM51}    & 
\colhead{0.5--8 keV}    &  
\colhead{0.5--2 keV}    &  
\colhead{2--8 keV}    &  
\colhead{0.5--8 keV}    &  
\colhead{0.5--2 keV}    &  
\colhead{2--8 keV}    &  
\colhead{0.5--8 keV}    &  
\colhead{0.5--2 keV}    &  
\colhead{2--8 keV}    \\ 
\colhead{(1)}	&
\colhead{(2)}	&
\colhead{(3)}	&
\colhead{(4)}	&
\colhead{(5)}	&
\colhead{(6)}	&
\colhead{(7)}	&
\colhead{(8)}	&
\colhead{(9)}	&
\colhead{(10)}	&
\colhead{(11)}	&
\colhead{(12)}	&
\colhead{(13)}	}
\startdata
1  &  13 29 26.63  &  47 12 34.5  &  J132926.6+471235  &  9.7$\pm$4.8  &  $<$14.9  &  $<$7.3  &  $<$10.8  &  $<$10.4  &  $<$4.1 
 &  $<$11.7  &  $<$8.3  &  $<$6.9  \\
2  &  13 29 33.33  &  47 14 40.4  &  J132933.3+471440  &  18.7$\pm$5.8  &  10.3$\pm$4.4  &  $<$16.3  &  $<$15.7  &  $<$12.2  &  $<$7.5 
 &  10.2$\pm$4.8  &  4.4$\pm$3.4  &  $<$12.7  \\
3  &  13 29 36.77  &  47 14 48.5  &  J132936.8+471448  &  22.0$\pm$6.3  &  20.7$\pm$5.9  &  $<$6.7  &  12.8$\pm$5.0  &  $<$19.2  &  $<$6.6 
 &  8.3$\pm$4.5  &  9.7$\pm$4.5  &  $<$4.0  \\
4  &  13 29 37.06  &  47 13 30.0  &  J132937.1+471330  &  33.6$\pm$7.2  &  20.2$\pm$5.8  &  13.5$\pm$5.0  &  $<$19.0  &  $<$13.2  &  $<$10.1 
 &  22.8$\pm$6.1  &  13.6$\pm$3.7  &  $<$16.8  \\
5  &  13 29 38.01  &  47 16 13.4  &  J132938.0+471613  &  44.7$\pm$8.0  &  39.9$\pm$7.5  &  $<$11.1  &  8.9$\pm$3.0  &  8.9$\pm$3.0  &  $<$4.5 
 &  36.1$\pm$7.3  &  30.9$\pm$6.8  &  $<$11.5  \\
6  &  13 29 38.63  &  47 13 36.2  &  J132938.6+471336  &  92$\pm$11  &  82.0$\pm$9.1  &  $<$17.8  &  20.1$\pm$5.7  &  19.2$\pm$5.6  &  $<$4.6 
 &  71.8$\pm$9.8  &  63.1$\pm$9.1  &  8.8$\pm$4.5  \\
7  &  13 29 42.18  &  47 14 48.1  &  J132942.2+471448  &  28.3$\pm$6.6  &  16.2$\pm$5.2  &  12.1$\pm$4.7  &  8.8$\pm$3.0  &  5.9$\pm$2.4  &  $<$7.4 
 &  19.9$\pm$5.7  &  10.2$\pm$4.4  &  $<$17.2  \\
8  &  13 29 43.36  &  47 15 25.3  &  J132943.4+471525  &  133$\pm$13  &  110$\pm$12  &  22.8$\pm$6.2  &  32.0$\pm$6.9  &  26.2$\pm$6.3  &  $<$12.4 
 &  101$\pm$11  &  84$\pm$10  &  17.0$\pm$5.5  \\
9  &  13 29 43.83  &  47 14 34.6  &  J132943.8+471435  &  7.5$\pm$2.8  &  $<$11.1  &  $<$7.9  &  $<$10.1  &  $<$5.9  &  $<$7.6 
 &  $<$9.0  &  $<$8.7  &  $<$4.1  \\
10  &  13 29 56.25  &  47 14 51.6  &  J132956.3+471452  &  161$\pm$14  &  119$\pm$12  &  41.8$\pm$7.7  &  $<$4.0  &  $<$4.3  &  $<$4.3 
 &  162$\pm$14  &  120$\pm$12  &  42.4$\pm$7.7  \\
11  &  13 30 06.47  &  47 08 34.3  &  J133006.5+470834  &  265$\pm$16  &  212$\pm$16  &  53.8$\pm$8.5  &  98$\pm$11  &  78.8$\pm$9.9  &  19.6$\pm$5.6 
 &  168$\pm$14  &  134$\pm$13  &  34.2$\pm$7.0  \\
12  &  13 30 15.18  &  47 15 39.0  &  J133015.2+471539  &  20.7$\pm$6.2  &  10.3$\pm$4.6  &  $<$19.5  &  $<$12.2  &  $<$7.2  &  $<$8.6 
 &  15.1$\pm$5.5  &  8.0$\pm$4.1  &  7.1$\pm$4.3  \\
13  &  13 30 15.74  &  47 15 16.5  &  J133015.7+471517  &  139$\pm$12  &  118$\pm$12  &  19.8$\pm$4.8  &  21.1$\pm$5.9  &  17.4$\pm$5.3  &  $<$9.4 
 &  116$\pm$12  &  101$\pm$11  &  16.1$\pm$4.2  \\
14  &  13 30 21.01  &  47 13 53.5  &  J133021.0+471354  &  32.9$\pm$7.3  &  29.0$\pm$5.6  &  $<$10.1  &  $<$14.4  &  $<$11.4  &  $<$6.8 
 &  25.8$\pm$6.5  &  23.9$\pm$6.1  &  $<$7.2  \\
15  &  13 30 25.11  &  47 13 13.2  &  J133025.1+471313  &  26.3$\pm$6.4  &  18.0$\pm$4.4  &  $<$16.8  &  18.7$\pm$5.5  &  12.7$\pm$4.8  &  $<$12.0 
 &  7.6$\pm$4.2  &  $<$10.3  &  $<$8.5  \\
16  &  13 30 28.45  &  47 15 20.3  &  J133028.5+471520  &  17.7$\pm$5.9  &  14.2$\pm$5.2  &  $<$9.3  &  $<$4.5  &  $<$4.5  &  $<$4.5 
 &  17.1$\pm$5.7  &  13.2$\pm$5.1  &  $<$9.7  \\
17  &  13 30 29.34  &  47 14 28.2  &  J133029.3+471428  &  19.1$\pm$6.0  &  13.9$\pm$3.9  &  $<$12.1  &  11.8$\pm$5.0  &  $<$13.5  &  $<$11.5 
 &  7.3$\pm$4.1  &  6.6$\pm$2.6  &  $<$4.2  \\
\enddata
\end{deluxetable}

\setcounter{table}{4}

\begin{center}

\begin{deluxetable}{ll}
\tablewidth{14cm}
	\tablecaption{ROSAT Source ID}
%\tabletypesize{\scriptsize}
\tablehead{
\colhead{CXO Name}	&
\colhead{ROSAT Source Name} \\
\colhead{(1)} &
\colhead{(2)}
}
\startdata
NGC 5194:\\
132939.5+471244 (5), 132940.0+471237 (6)	& E4, RW1, M7, IXO78, R1\\
132943.3+471135	(9)				& E5, RW2, M6, IXO79, R2\\
132945.9+471056	(16)				& M5, R3\\
132946.1+471042	(17)				& E6, RW3\\
...						& E7, M4, R4\\
132950.8+471031 (27), 132951.4+471032 (28)	& E8\\
132953.7+471436	(41), 132953.8+471432 (42)	& E10, RW5, M8, R6\\
133001.0+471344 (69), 133001.1+471333 (70)	& E12, RW7, M1, IXO80, R8\\
133004.3+471321 (76)				& M2\\
133007.6+471106 (82)				& E14, RW9, M3, IXO81, R9\\
Nuclear region					& E9, RW4, R5\\
NGC 5195:\\
132959.5+471559	(NGC 5195 - 7) 			& E11, RW6, R7\\
133006.0+471542 (NGC 5195 - 12)			& E13, NGC 5195 RW1\\
133006.5+470834 (outside NGC 5194/95)		& RW8\\
Outside {\it Chandra} FOV			& E1, E2, E3, E15, E16\\
\enddata
\tablecomments{Col. (1): {\it Chandra} name CXOM51~J. The source numbers in
Tables 2 and 3 are shown in the parentheses. Col. (2): {\it ROSAT} source name.
E, M, RW, IXO, and R denote the sources in Ehle et al. (1995), Marston et al. (1995),
Roberts \& Warwick (2000), Colbert \& Ptak (2002), and Read, Ponman, \& Strickland (1997), 
respectively.
}
\end{deluxetable}

%%%%%% Spectral fits (Tables 6 & 7) %%%%%%

%Luminous sources

%L(0.5-8 keV) $>$1e39 in at least one obs.

\setcounter{table}{5}

\begin{deluxetable}{llccccccc}
%\rotate
\tablewidth{15cm}
	\tablecaption{Spectral Models of Ultraluminous Sources
($L_{0.5-8 \rm keV}>10^{39}$ ergs s$^{-1}$)}
\tabletypesize{\scriptsize}
\tablehead{
\colhead{CXO Name}     &
\colhead{No.}     &
\colhead{$N_{\rm H}$} & 
Photon Index $\Gamma$ & 
$kT_{\rm in}$ (MCD) & 
$\chi^2$/dof    & 
$\chi^2$/dof    & 
Flux & 
Luminosity \\
\colhead{CXOM51}	&
	&
($10^{22}$ cm$^{-2}$) &
	&
(keV)	&
(Power law)	&
(MCD)	&
(0.5$-$8 keV)	&
(0.5$-$8 keV)	
\\
\colhead{(1)}	&
\colhead{(2)}	&
\colhead{(3)}	&
\colhead{(4)}	&
\colhead{(5)}	&
\colhead{(6)}	&
\colhead{(7)}	&
\colhead{(8)}	&
\colhead{(9)}
}
\startdata
NGC 5194:\\
J132939.5+471244 (1)& 5 	& 0.18 (0.074-0.29)	& 1.37 (1.09-1.76) & ...& 11.8/12 &...	& 1.61 	& 1.52 \\
	& 	& 0.099 (0.030-0.20)	& ...& 1.58 (1.11-2.70)	& ... & 10.6/12	& 1.27 	& 1.15 \\
J132939.5+471244 (2)& 5 	& 0 ($<$0.087)		& 1.22 (0.98-1.49) & ...& 11.4/8 &...	& 0.63	& 0.54\\
	& 	& 0 ($<0.043$)		& ...& 1.37 (0.98-2.16)	& ...&11.9/9	& 0.44	& 0.38\\
%src103	& 5	& \\
J132943.3+471135 (1)& 9 	& 0.7			& 10	& ...		& 20.4/10&... 	& ...	& ...	\\ % & ultrasoft\\
	&	& 0.40 (0.25--0.50)	& ...	& 0.084 (0.079-0.090)	&...& 7.0/9	& 0.27 & 2.74  \\
J132943.3+471135 (2)& 9 	& 0.6			& 8.2	& ...		& 40.0/14&...	& ...	& ...	\\
	&	& 0.20 (0.13--0.38)	& ...	& 0.12 (0.096-0.14)	&...& 20.5/14	& 0.21 & 0.62 \\
%src98	& 9	& 0.29 (0.20-0.45)	& ...& 0.087 (0.074-0.098) & 17.4/16	& 0.23 & 1.10  & Blackbody fits\\
J132950.7+471155 (1)	& 26 	& 5.8 (0.99-12.7)	& 1.75 (0.73-3.85)	&...& 3.2/3&...	& 1.85 	& 2.95 \\ % emission lines\\
J132950.7+471155 (2)	& 26 	& 3.6 (2.9-5.6)		& 1.55 (0.95-2.41)	& ...& 12.2/14&...	& 2.32 	& 3.95 \\
J132953.3+471042 (2)	& 37 	& 0.11 (0.060-0.15)	& 1.55 (1.40-1.74) & ... & 25.0/33&...	& 1.58 	& 1.47 \\
	& 	& 0 ($<$0.023)		& ...& 1.76 (1.39-2.05)	&...& 35.9/33	& 1.49 	& 1.26 \\
J132953.7+471436 (1)	& 41 	& 0.11 ($<$0.18)	& 1.50 (1.28-1.88) & ...& 6.8/11 &... 	& 1.21	& 1.13 \\
	& 	& 0.018 ($<$0.11)	& ...	& 1.43 (1.03-2.10) &...& 5.5/11 	& 0.98	& 0.84\\
J132953.7+471436 (2)	& 41 	& 0.090 (0.033-0.14)	& 1.32 (1.15-1.49) & ...& 27.1/23 &...	& 1.26	& 1.13 \\
	& 	& 0.0012 ($<$0.042)	& ...	& 2.20 (1.71-2.84) &...& 26.6/23	& 1.20 	& 1.02 \\
%src63	& 41	& \\
J133001.0+471344 (1)	& 69 	& 0.12 (0.062-0.15)	& 1.24 (1.07-1.36) & ...& 34.9/28&...	& 2.99 	& 2.71 \\
	& 	& 0.034 ($<$0.073)	& ...	& 2.34 (1.81-3.31) &...& 37.0/28	& 2.79 	& 2.40 \\
J133001.0+471344 (2)	& 69 	& 1.6 (0.70-2.0)	& 9.4 ($>5.1$)	& ...	& 3.6/6	$^a$	&...& 0.04	& 57 \\
	&	& 0.82 (0.32-1.6)	& ...	 & 0.17 (0.11-0.30)&...& 3.3/6 $^a$	& 0.04	& 0.56 \\
J133007.6+471106 (1)	& 82 	& 0.16 (0.11-0.21) & 2.26 (2.05-2.50)	& ...	& 52.7/47 &...	& 2.85	& 3.14 \\
	&	& 0 ($<$0.021)	& ...	& 0.87 (0.78-0.96)	&...& 60.5/47 	& 2.46	& 2.08 \\
J133007.6+471106 (2)	& 82 	& 0.097 (0.056-2.0) & 1.86 (1.69-2.05)	& ...	& 66.0/47 &...	& 1.88 	& 1.80 \\
	& 	& 0		& ...			& 1.11	&...& 91.9/47	& ...	& ...\\
\\
NGC 5195:\\
J133006.0+471542 (1)	& 12 	& 0.10 ($<$0.19)	& 1.42 (1.22-1.82)	& ... & 17.3/11&... & 1.10 	& 1.01  \\
	&	& 0 ($<$0.092) 		& ...& 1.54 (1.11-2.27)	&...& 14.9/11	& 0.91	& 0.79 \\
J133006.0+471542 (2)	& 12 	& 0.11 (0.046-0.17)	& 1.32 (1.11-1.51)	& ... & 19.9/19&...& 1.20 	& 1.09  \\
	&	& 0.039 ($<$0.086)	& ...& 1.78 (1.32-2.64)	&...& 19.0/19	& 1.01 	& 0.87 \\
J132958.4+471547 (2)	& 5 	& 1.0	& 10	& ...				& 22.7/11&...	& ...	& ... \\ % & ultra soft\\
	&	& 0.80 (0.18-1.01)	& ... & 0.102 (0.058-0.227)&...	& 8.9/5	& 0.062& 2.74  \\
\enddata
\tablecomments{Col. (1): CXO Name (CXOM51). (1) and (2) denote
the first (June 2000)  and second (June 2001) observation, respectively.
Col. (2): Source number in Table 2 (NGC 5194, upper part of Table) and Table 3 (NGC 5195, lower part of Table). 
Col. (3): Absorbing column density. 
Col. (4): Photon index for power-law model.
Col. (5): Inner disk temperature ($kT_{\rm in}$) for multi-color disk model.
Col. (6): $\chi^2$ / dof for power law model.
Col. (7): $\chi^2$ / dof for MCD model.
Col. (8): Observed flux (not corrected for absorption) in the 0.5--8 keV band in units of 
$10^{-13}$ erg s$^{-1}$ cm$^{-2}$.
Col. (9): Luminosity (corrected for absorption) in the 0.5--8 keV band in units of $10^{39}$ erg s$^{-1}$. (a) $C$-statistic.
}
\end{deluxetable}

\clearpage

\begin{deluxetable}{llccccccc}
%\rotate
\tablewidth{15cm}
	\tablecaption{Spectral Models of Sources with
$L_{0.5-8 \rm keV}<10^{39}$ ergs s$^{-1}$}
\tabletypesize{\scriptsize}
\tablehead{
\colhead{CXO Name}     &
\colhead{No.}     &
\colhead{$N_{\rm H}$} & 
Photon Index $\Gamma$  & 
$kT_{\rm in}$ (MCD) & 
$\chi^2$/dof    & 
$\chi^2$/dof    & 
Flux &
Luminosity\\
\colhead{CXOM51} &
	&
($10^{22}$ cm$^{-2}$) &
	&
(keV)	&
(Power law)	&
(MCD)	&
(0.5$-$8 keV)	&
(0.5$-$8 keV)	
\\
\colhead{(1)}	&
\colhead{(2)}	&
\colhead{(3)}	&
\colhead{(4)}	&
\colhead{(5)}	&
\colhead{(6)}	&
\colhead{(7)}	&
\colhead{(8)}	&
\colhead{(9)}
}
\startdata
NGC 5194:\\
J132940.0+471237 (1)& 6 	& 0.41 (0.26-0.62)	& 4.47 (3.49-5.88)	&...& 11.6/13&...	& 0.48	& 1.96  \\
	&	& 0.14 (0.051-0.23)	& ...& 0.30 (0.24-0.41)	&...& 13.7/13	& 0.43	& 0.59 \\ 
%src102-2& 6(2)	& 			&	&		&	& & & too faint to fit\\
%src102	& 6	& \\
J132949.0+471053 (2)	& 21 	& 0.057 ($<$0.16)	& 1.26 (0.80-1.69)	& ...& 3.8/5&...	& 0.47 & 0.42 \\
	&	& 0 ($<$0.13)	& ...& 1.75 (1.02-4.38)		&...& 4.2/5		& 0.38 & 0.32 \\
J132950.8+471031 (1)	& 27 	& 0.50 (0.27-0.99) & 3.6 (2.6-6.2)	& ...	& 1.3/3	&...	& 0.18 & 0.58 \\
	& 	& 0.22 ($<$0.54)& ...	& 0.43 (0.25-0.97)	&...& 1.6/3		& 0.15 & 0.21 \\
J132951.4+471032 (2)	& 28 	& 0.19 (0.069-0.31)	& 1.44 (1.13-1.83) & ...& 3.4/7	&...	& 0.54 & 0.52 \\
	&	& 0.086 ($<$0.21)	& ...	& 1.72 (1.19-2.85)	&...& 3.1/7	& 0.47 & 0.42 \\
%src77	& 28	& \\
J132953.8+471432 (1)& 42 	& 0.23 ($<$0.46)	& 2.37 (2.00-3.15)	& ...& 1.18/4&...	& 0.41 & 0.51 \\
	&	& 0 ($<$0.16)	& ...	& 0.90 (0.61-1.20)	&...& 2.0/4		& 0.37 & 0.31	\\
J132953.8+471432 (2)	& 42 	& 0.17 (0.037-0.30)	& 1.91 (1.51-2.25)	& ... & 5.8/8&...	& 0.39 & 0.40 \\
	&	& 0 ($<$0.14)	& ...	& 1.02 (0.64-1.71)	&...& 9.0/7		& 0.30 & 0.26	\\
%src62	& 42	& \\
J132957.6+471048 (1)	& 63 	& 0.24 (0.061-0.42)	& 1.88 (1.49-2.48)	& ...& 4.6/5&...	& 0.57 & 0.62 \\
	&	& 0.10 ($<$0.24)	& ...& 1.06 (0.71-1.96)	&...& 4.8/5		& 0.45 & 0.42 \\ 
J132957.6+471048 (2)	& 63 	& 0.26 (0.15-0.36)	& 2.13 (1.82-2.53)	& ...& 11.0/15&...	& 0.64 & 0.76 \\
	&	& 0.094 (0.029-0.17)	& ...& 0.91 (0.72-1.28)	&...& 10.0/15	& 0.52 & 0.49 \\
%src37	& 63	& \\
J132957.6+471206 (1)	& 64 	& 0 ($<$0.40)	& 0.98 (0.36-2.82)	& ...	& 0.83/2&...	& 0.91 & 0.77 \\
	&	& 0 ($<$0.27)	& ...	& 1.93 		&...& 0.85/2	& 0.63 & 0.53 \\ %& $kT$ not constrained\\
J133004.3+471321 (2)	& 76 	& 0 ($<$0.16)	& 1.19 (0.84-1.64)	& ...	& 7.0/4&...		& 0.37 & 0.32 \\
	&	& 0 		& ...			& 1.86	&...& 8.8/4		& ...	& ... \\
\\
NGC 5195:\\
J132952.8+471644 (2)	& 1 	& 0 ($<$0.14)	& 1.03 (0.68-1.45) &		& 4.2/3&...		& 0.37 & 0.31 \\
	&	& 0 ($<$0.11)	& ...	& 2.36 ($>$1.51) 	&...& 5.3/3 	& ...	& ...	\\
%src68	&	& \\
J132957.5+471612 (1)	& 4 	& 0.0058 ($<$0.33)	& 1.00 (0.68-1.62)	& ...& 0.40/4&...	& 0.71 & 0.60 \\
	&	& 0 ($<$0.11)	& ...	& 2.70 ($>$1.58)	&...& 0.74/4	& 0.63 & 0.54 \\
J132957.5+471612 (2)	& 4 	& 0.50 (0.27-0.79)	& 1.85 (1.37-2.81)	& ...& 13.4/8&...	& 0.48 & 0.59 \\
	&	& 0.25 (0.055-0.59)	& ...& 1.31 (0.83-2.56)	&...& 13.3/8	& 0.40 & 0.41 \\
%src38	&	& \\							
J132958.7+471625 (2)	& 6 	& 0.41 (0.051-0.81)& 3.3 (2.6-5.1)	& ...	& 3.8/4	&...	& 0.14 & 0.34 \\
	&	& 0.14 ($<$0.48)& ...	& 0.50 (0.33-0.84)	&...& 3.4/4		& 0.13 & 0.14 \\
J132959.5+471559 (1)	& 7 	& 0 ($<$0.087)	& 1.43 (1.18-1.75)	&	& 13.6/14 &...	& 0.88 & 0.74 \\
	&	& 0 ($<$0.041)	& ...	& 1.59 (1.13-2.34)	&...& 19.0/14 	& 0.77 & 0.66 \\
J132959.5+471559 (2)	& 7 	& 0.11(0.234-0.19)	& 1.80 (1.58-2.06) & ...& 25.0/19&...	& 0.64 & 0.80 \\
	&	& 0 ($<$0.047)	& ...	& 1.22 (1.01-1.45)	&...& 26.9/19 	& 0.74 & 0.63 \\
%src30	& 7	& 0.11		& 1.69	&	& 41.6/21	& \\				
J133003.8+471611 (2)	& 11 	& 0.43 (0.003-1.1) & 3.3 (1.7-6.3)	& ... 	& 3.2/3&...		& 0.13 & 0.31 \\
	&	& 0.20 ($<$0.67)& ...	& 0.45 (0.26-1.04)	&...& 2.78/3	& 0.10 & 0.13 \\
\\
Outside NGC 5194/5195: $^a$\\									
J132943.4+471525 (2)	& 8 	& 0.099 ($<0.23$)& 2.02 (1.19-3.43)	& ...	& 0.76/3&...	& 0.22 & ...\\ %0.213 \\
	&	& 0 ($<$0.16)	& ...	& 0.74 (0.45-1.17)	&...& 0.69/3	& 0.15 & ...\\ %0.129 \\
J132956.3+471452 (2)	& 10  	& 0.17 ($<$0.29)	& 1.87 (1.25-2.23)	& ...& 2.3/7&...	& 0.44 & ...\\ %0.455 \\
	&	& 0 ($<$0.16)	& ... & 1.25 (0.81-1.65)	&...& 3.0/7		& 0.39 & ...\\ %0.331 \\
J133006.5+470834 (1)	& 11 	& 0 ($<$0.17)	& 1.75 (1.41-2.43)	& ...	& 2.18/3&...& 0.43 & ...\\ %	\\
	&	& 0 ($<$0.099)	& ...& 1.00 (0.69-1.40)		&...& 4.4/3		& 0.34 & ...\\ %	\\
J133006.5+470834 (2)	& 11 	& 0.17 (0.028-0.28)	& 1.88 (1.57-2.43)	& ...& 8.5/8&...	& 0.42 & ...\\ %0.426 \\
	&	& 0.0071 ($<$0.11)	& ...& 1.22 (0.88-1.61)	&...& 8.3/8		& 0.37 & ...\\ %0.312 \\
%src10	&	& \\
J133015.7+471517 (2)	& 13 	& 0.14		& 2.2			& ...	& 14.1/6&...	& 0.21 & ...\\ %0.223 \\
	&	& 0.009		& ...& 0.75			&...& 12.7/6 	& 0.17 & ...\\ %0.141 \\
\enddata
\tablecomments{Col. (1): CXO Name (CXOM51). (1) and (2) denote
the first (June 2000) and second (June 2001) observation, respectively.
Col. (2): Source number in Table 2 (NGC 5194, upper part of Table), Table 3 (NGC 5195, middle part of Table), and Table 4 (spatially outside NGC 5194 and NGC 5195, bottom part of Table). 
Col. (3): Absorbing column density. 
Col. (4): Photon index for power-law model.
Col. (5): Inner disk temperature ($kT_{\rm in}$) for multi-color disk model.
Col. (6): $\chi^2$ / dof for a power law model.
Col. (7): $\chi^2$ / dof for a MCD model.
Col. (8): Observed flux (not corrected for absorption) in the 0.5--8 keV band in units of 
$10^{-13}$ erg s$^{-1}$ cm$^{-2}$.
Col. (9): Luminosity (corrected for absorption) in the 0.5--8 keV band in units of $10^{39}$ erg s$^{-1}$.
(a) Luminosities are $<10^{39}$ erg s$^{-1}$ if these sources are associated with M51.
}
\end{deluxetable}

\setcounter{table}{7}

\begin{deluxetable}{lccccc}
\tablewidth{17cm}
	\tablecaption{Spectral models of NGC 5194 \#26}
%\tabletypesize{\scriptsize}
\tablehead{
\colhead{}	&
\colhead{Photon index$^a$}	&
\colhead{$N_{\rm H}$}	&	
\multicolumn{2}{c}{Gaussians$^b$}	&
\colhead{Suggested ID}	\\
\colhead{}	&
\colhead{}	&
\colhead{($10^{22}$ cm$^{-2}$)}	&
\colhead{Energy (keV)}&
\colhead{EW (eV)} &
\colhead{}\\
}
\startdata
Observation 1:	& 1.55 (1.41-1.70) & ...\\
		& ...	& 4.6 (3.8--5.5)\\
%		& Gaussians	& Energy (keV)		& EW (eV)\\
		&...&...& 1.80 (1.52--1.91)	& 187 ($<$507)		& \ion{Si}{2} -- \ion{Si}{13} K$\alpha$ \\
		&...&...& 3.24 (3.18--3.29)	& 386 (140--744)	& \ion{Ar}{18} K$\alpha$?\\
		&...&...& 4.03 (3.96--4.10)	& 371 (130--694)	& \ion{Ca}{20} K$\alpha$\\
		&...&...& 6.65 (6.50--6.74)	& 1070 (217-2420)	& \ion{Fe}{20} -- \ion{Fe}{25} K$\alpha$\\
Observation 2:	& 1.78 (1.38--2.48)	& ...\\
		& ... & 0.11 ($<$0.43)\\
		& ... & 4.5 (3.5--5.1) & (CF$^c$ 0.988 (0.973--0.995))\\
\enddata
\tablecomments{
(a) The photon indices are slightly different from those in Table 6 because, in that table,
they are derived assuming a simple power law model
without Gaussians and partial covering.
(b) Centroid energies of Gaussians are as observed. Lines are 
assumed to be narrow.
(c) Covering factor for partial covering model.
}
\end{deluxetable}

\begin{deluxetable}{llccccccc}
\tablewidth{17cm}
        \tablecaption{Spectral Models of Soft Sources}
%\tabletypesize{\scriptsize}
\tablehead{
\colhead{CXO Name}     &
\colhead{No.}     &
\colhead{$N_{\rm H}$} & 
\colhead{$kT$} & 
\colhead{Abundance} & 
\colhead{$\chi^2$/dof}    & 
\colhead{Flux} & 
\colhead{Luminosity}  \\
\colhead{CXOM51}        &
        &
\colhead{($10^{22}$ cm$^{-2}$)} &
\colhead{(keV)}   &
\colhead{(solar)}	& 
        &
\colhead{(0.5$-$8 keV)}   &
\colhead{(0.5$-$8 keV)}   &
\\
\colhead{(1)}     &
\colhead{(2)}     &
\colhead{(3)}     &
\colhead{(4)}     &
\colhead{(5)}     &
\colhead{(6)}     &
\colhead{(7)}	&
\colhead{(8)}
}
\startdata
Blackbody\\
J132943.3+471135 (1) & NGC 5194 9	&$0.26^{+0.39}_{-0.24}$ 	& $0.084^{+0.029}_{-0.025}$	& ...	& 6.7/10	& 0.27	& 2.6 (0.79--14)\\
J132943.3+471135 (2) & NGC 5194 9	& $0.15^{+0.15}_{-0.07}$	& $0.11\pm0.02$		& ...	& 18.5/14	& 0.12	& 0.40 (0.26 -- 1.0)\\
J132958.4+471547 (2) & NGC 5195 5	& $0.67^{+0.60}_{-0.66}$	& $0.099^{+0.092}_{-0.036}$	& ...	& 8.9/5	& 0.062	& 1.3 (0.056--25)\\
MEKAL plasma\\
J132943.3+471135 (1) & NGC 5194 9	& $0.44^{+0.24}_{-0.15}$	& $0.11\pm0.03$		& $<0.010$	& 8.6/9	& \\
J132943.3+471135 (2) & NGC 5194 9	& $0.070^{+0.13}_{-0.06}$	& $0.24^{+0.04}_{-0.06}$& $0.038^{+0.075}_{-0.028}$	& 16.7/13	& \\ %Fx = 0.22	
J132958.4+471547 (2) & NGC 5195 5	& 0.97				& 0.11			& 0.028		& 8.5/4\\
\enddata
\tablecomments{Col. (1): CXO name (CXOM51). (1) and (2) denote the first
(June 2000) and second (June 2001) observation, respectively. 
Col. (2): Source name in Table 2 (NGC 5194) and Table 3 (NGC 5195).
Col. (3): Absorbing column density. 
Col. (4): $kT$ for blackbody or MEKAL model.
Col. (5): Abundance for MEKAL model.
Col. (6): $\chi^2$ / dof.
Col. (7): Observed flux (not corrected for absorption) in the 0.5--8 keV band in unit of $10^{-13}$ erg s$^{-1}$ cm$^{-2}$.
Col. (8): Luminosity corrected for absorption in the 0.5--8 keV band in units of $10^{39}$ erg s$^{-1}$. The range of allowed luminosities is shown in the parentheses.
}
\end{deluxetable}

\end{center}

\clearpage

\begin{figure*}[h]
\begin{center}
\includegraphics[scale=0.65,angle=-90]{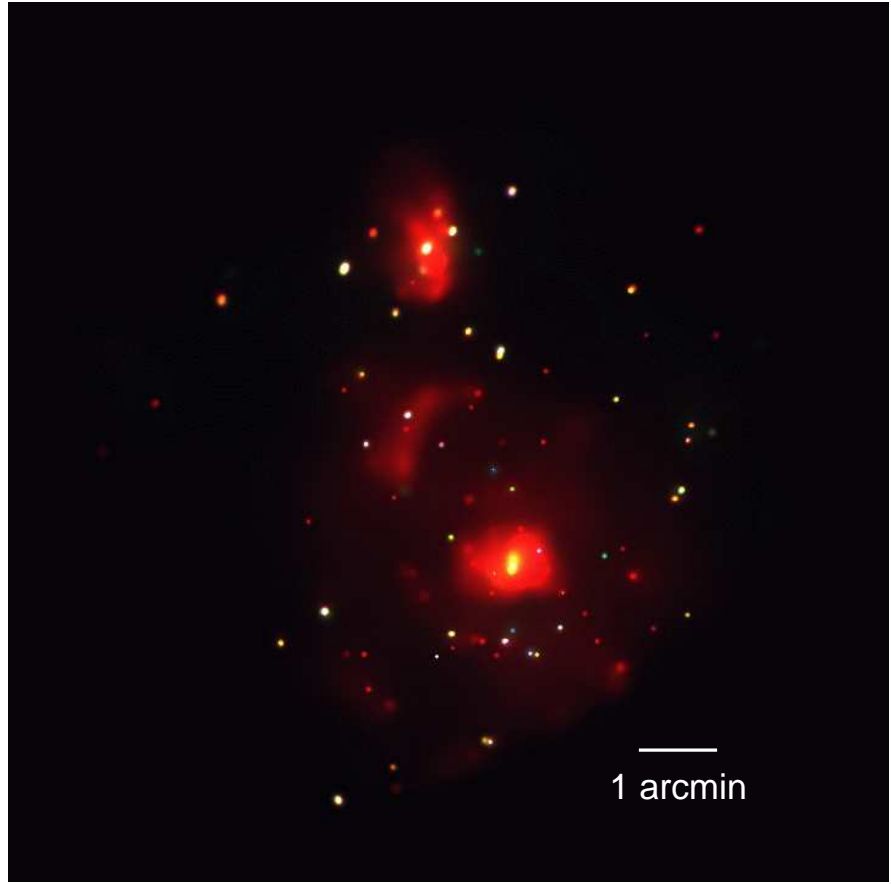}
\figcaption[]{
Adaptively smoothed color image of M51. Red: 0.3--1.5 keV, Green: 1.5--3 keV, and Blue: 3-8 keV. North is up and East is to the left.
}
\end{center}
\end{figure*}

\begin{figure*}[ht]
\begin{center}
\includegraphics[scale=0.8,angle=0]{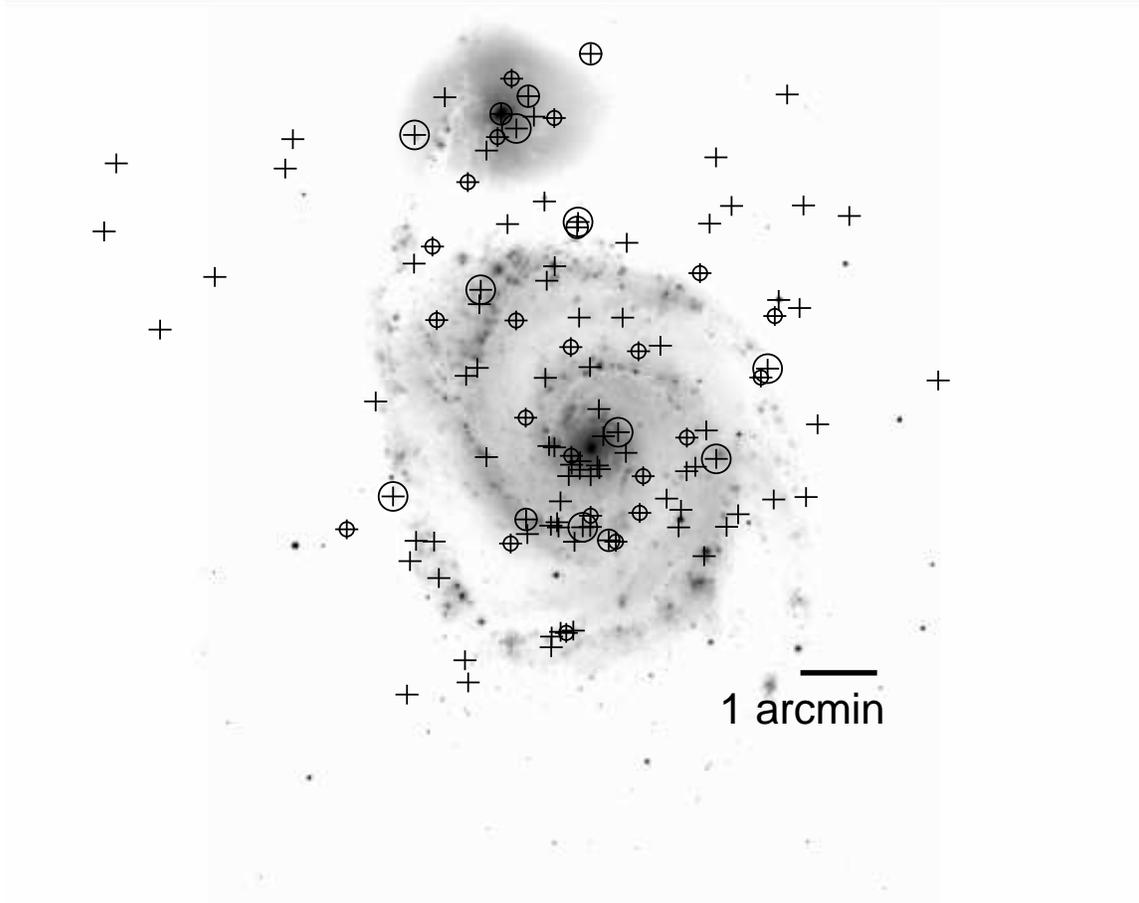}
\figcaption[]{
X-ray source positions (crosses) overlaid on a narrow
band (H$\alpha$, continuum not subtracted) optical image (Thilker et
al. 2000). Crosses surrounded by large, medium, and small circles 
correspond to sources with
$L$(0.5 -- 8 keV) $>$ 10$^{39}$ {\eps},
$L$(0.5 -- 8 keV) = (5 -- 10) $\times$ 10$^{38}$ {\eps}, and
$L$(0.5 -- 8 keV) = (1 -- 5) $\times$ 10$^{38}$ {\eps} in the 0.5 -- 8 keV 
band, respectively. Crosses without circles correspond to sources
with $L$(0.5 -- 8 keV) $\le$ 1$\times$ 10$^{38}$ {\eps}. 
North is up and East is to the left.
}
\end{center}
\end{figure*}

\begin{figure*}[t]
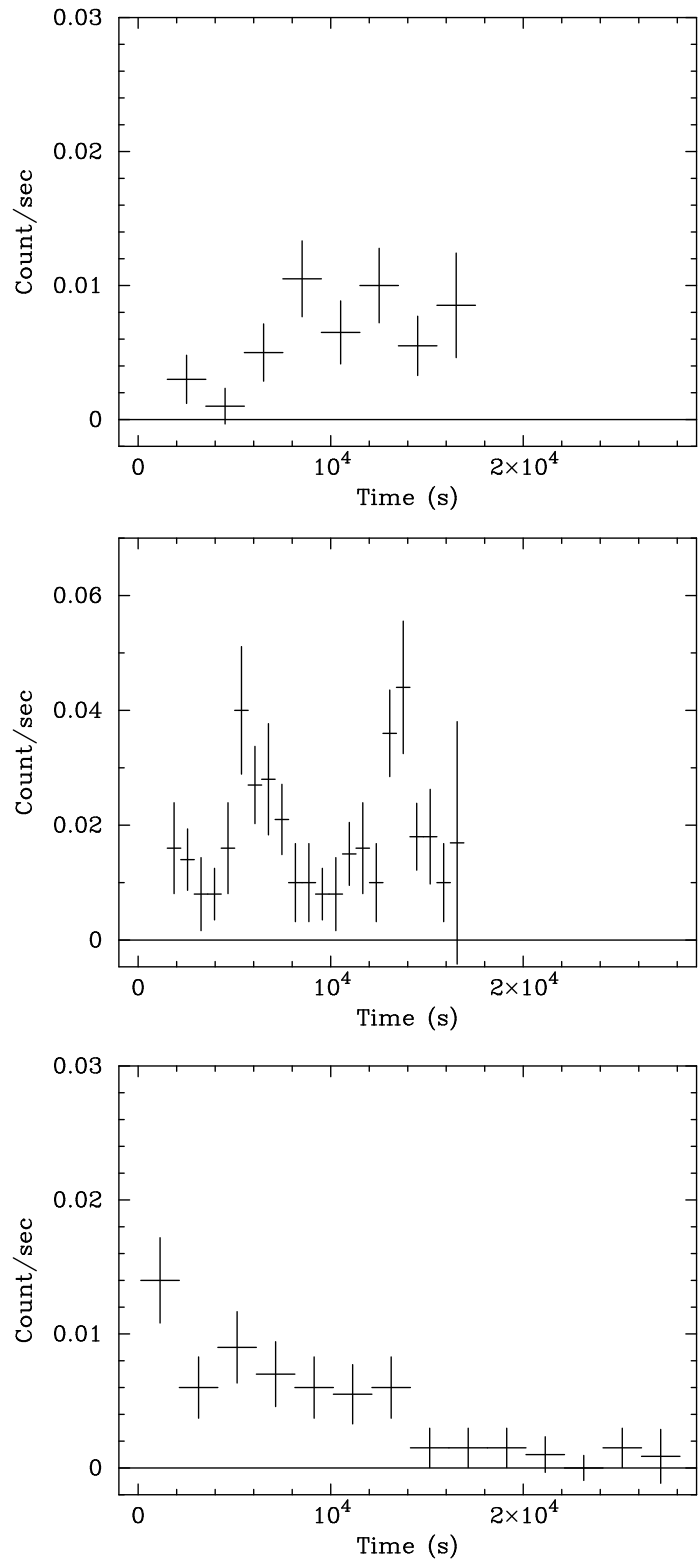

\begin{center}
\noindent
\includegraphics[scale=0.48,angle=-90]{f3a.ps}

\vspace{3mm}

\includegraphics[scale=0.48,angle=-90]{f3b.ps}

\vspace{3mm}

\includegraphics[scale=0.48,angle=-90]{f3c.ps}
\figcaption[]{
Light curves of bright X-ray sources.
The background is negligible and has not been subtracted. (a) NGC 5194
\#63 = CXOM51~J132957.6+471048 in year 2000 (bin size = 2000 sec), (b)
NGC 5194 \#69 = CXOM51~J133001.0+471344 in year 2000 (bin size = 700
sec), and (c) NGC 5194 \#76 = CXOM51~J133004.3+471321 in year 2001
(bin size = 2000 sec).
}
\end{center}
\end{figure*}

\begin{figure*}[ht]
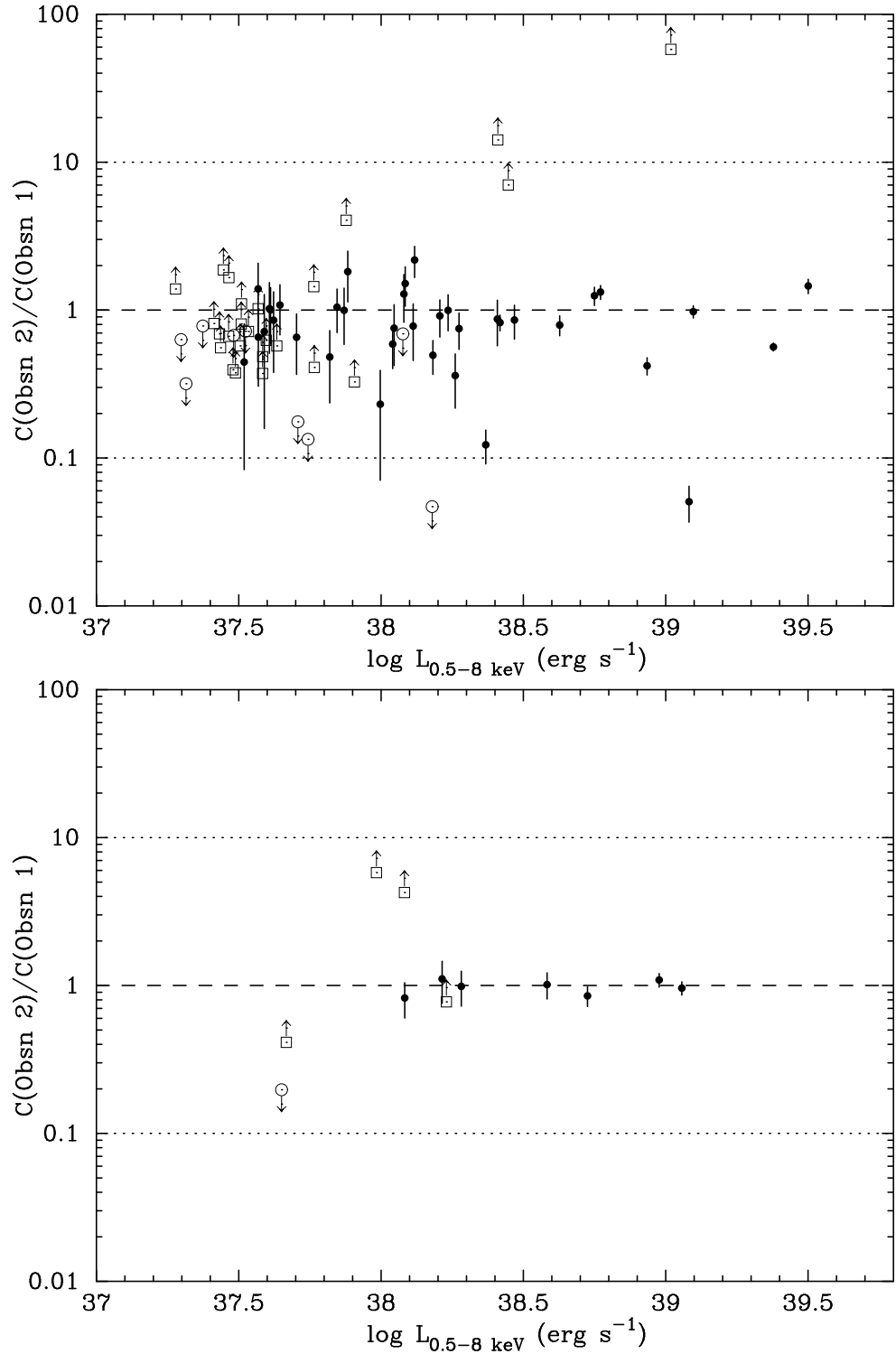

\begin{center}
\includegraphics[scale=0.55,angle=-90]{f4a.ps}

\includegraphics[scale=0.55,angle=-90]{f4b.ps}
\figcaption[]{
Ratios of the count rate (0.5--8 keV band) in the second (June 2001)
to that in the first (June 2000) observation as a function of the
luminosity in the 0.5--8 keV band. The errors are one-sigma.  Upper
and lower limits are represented by open circles and open squares,
respectively.  (a) Sources within NGC 5194. (b) Sources within NGC
5195.}
\end{center}
\end{figure*}

\begin{figure*}[ht]
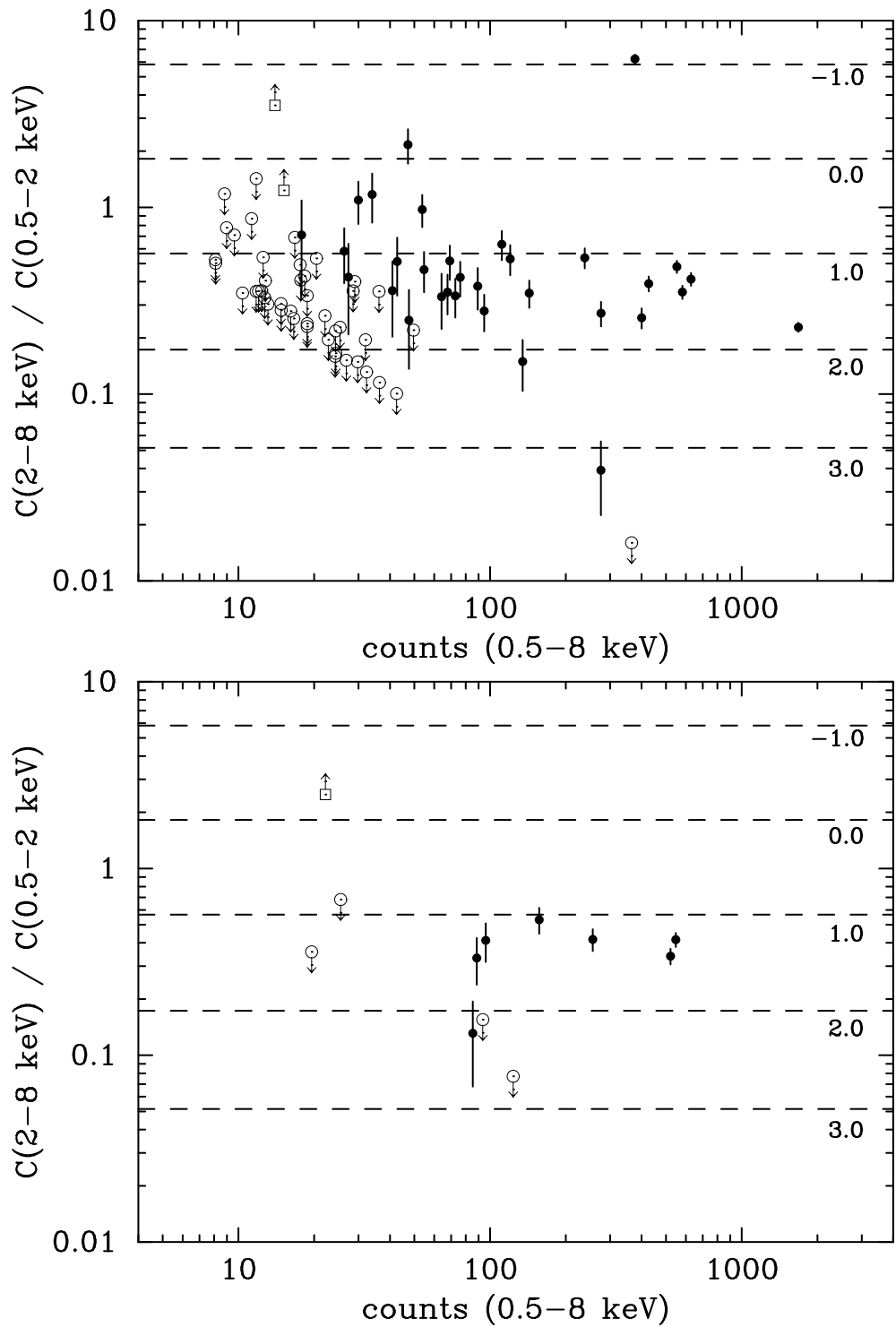

\begin{center}
\includegraphics[scale=0.7,angle=-90]{f5a.ps}

\includegraphics[scale=0.7,angle=-90]{f5b.ps}
\figcaption[]{
Dependence of the band ratio (count rate in 2--8 keV band / count rate
in 0.5--2 keV band) on counts in the 0.5--8 keV band for the detected
X-ray sources. The horizontal dashed lines correspond to photon
indices of --1, 0, 1, 2, and 3 when a power law model absorbed by the
Galactic column is assumed. The counts are as observed, and not
corrected for the effect of vignetting, and were obtained from the sum
of the year 2000 and 2001 observations. (a) Sources within NGC
5194. (b) Sources within NGC 5195.  }
\end{center}
\end{figure*}

\begin{figure*}[ht]
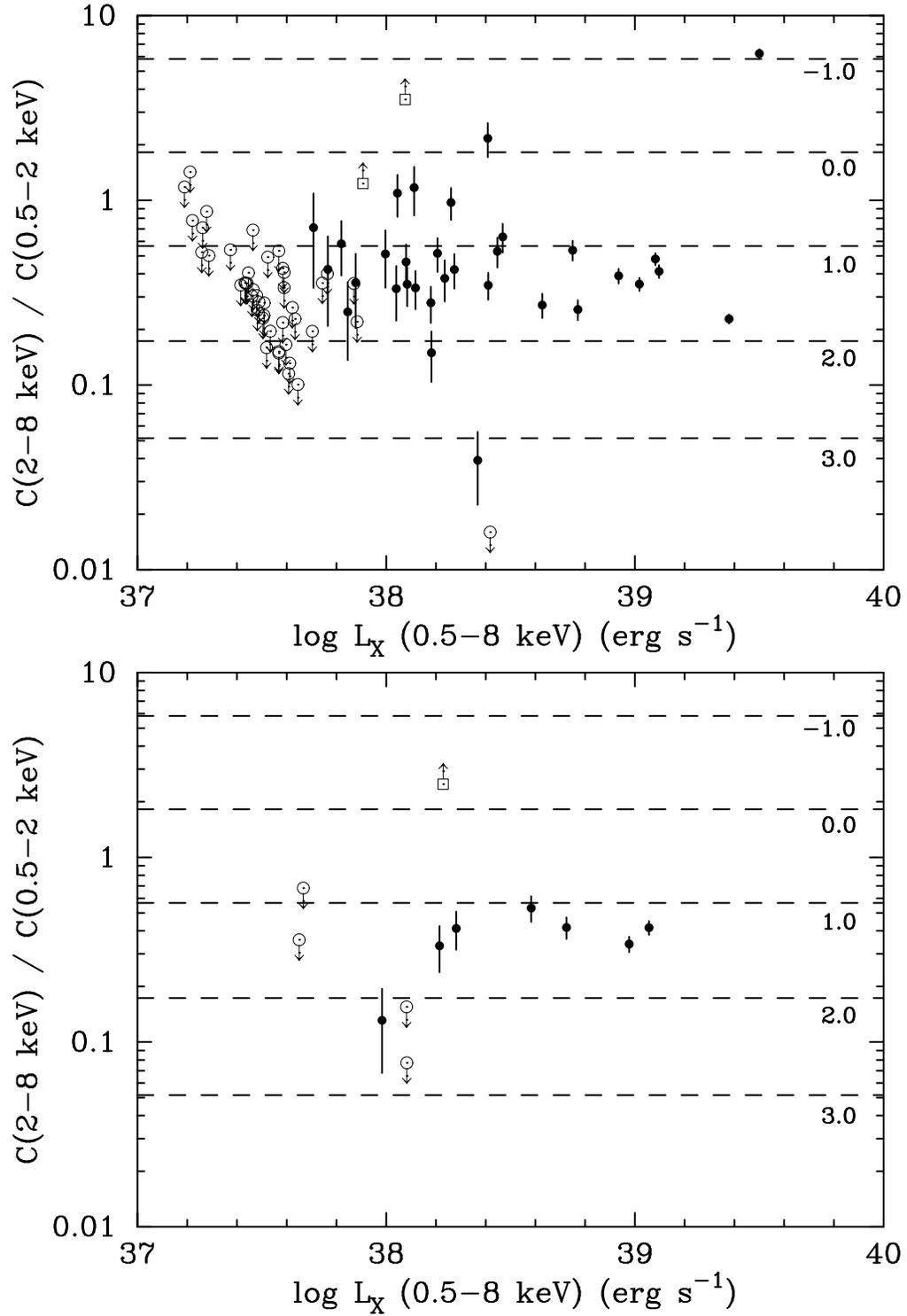

\begin{center}
\includegraphics[scale=0.7,angle=-90]{f6a.ps}

\includegraphics[scale=0.7,angle=-90]{f6b.ps}
\figcaption[]{
Dependence of the band ratio (count rate in 2--8 keV band / count rate
in 0.5--2 keV band) on luminosity in the 0.5--8 keV band for the
detected X-ray sources. The horizontal dashed lines correspond to
photon indices of --1, 0, 1, 2, and 3 when a power law model absorbed
by the Galactic column is assumed. The luminosities are corrected for
vignetting.  (a) Sources within NGC 5194. (b) Sources within NGC 5195.
}
\end{center}
\end{figure*}

\begin{figure*}[ht]
\begin{center}
\includegraphics[scale=0.7,angle=-90]{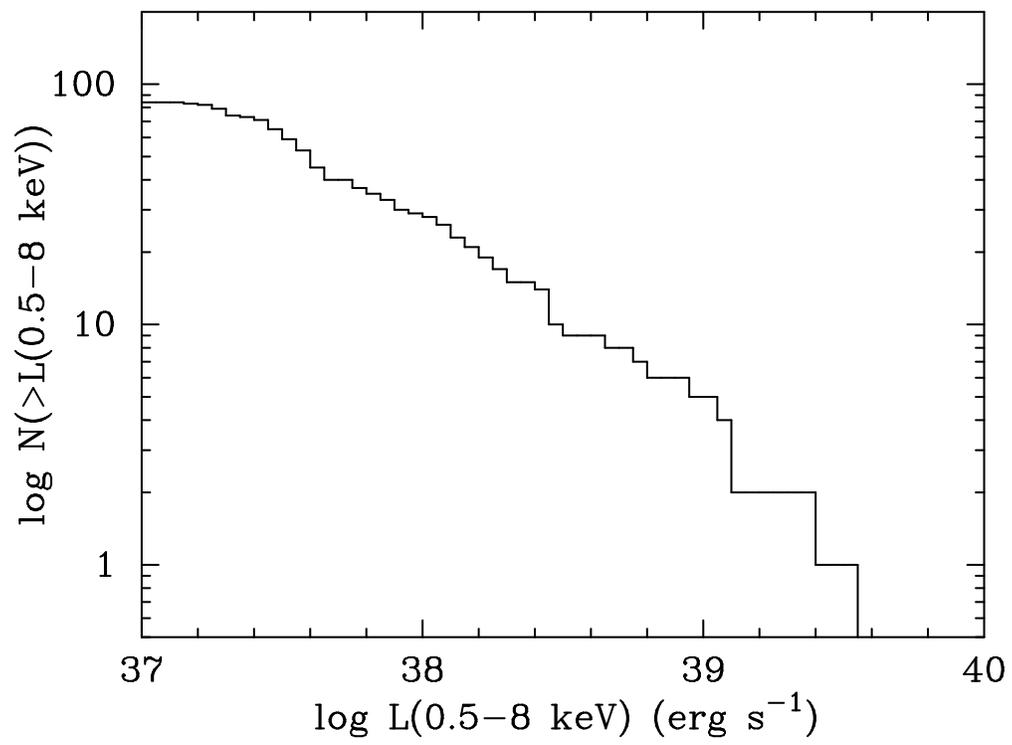}
\figcaption[]{
Cumulative luminosity function of the X-ray sources in NGC 5194.
}
\end{center}
\end{figure*}

\begin{figure*}[ht]
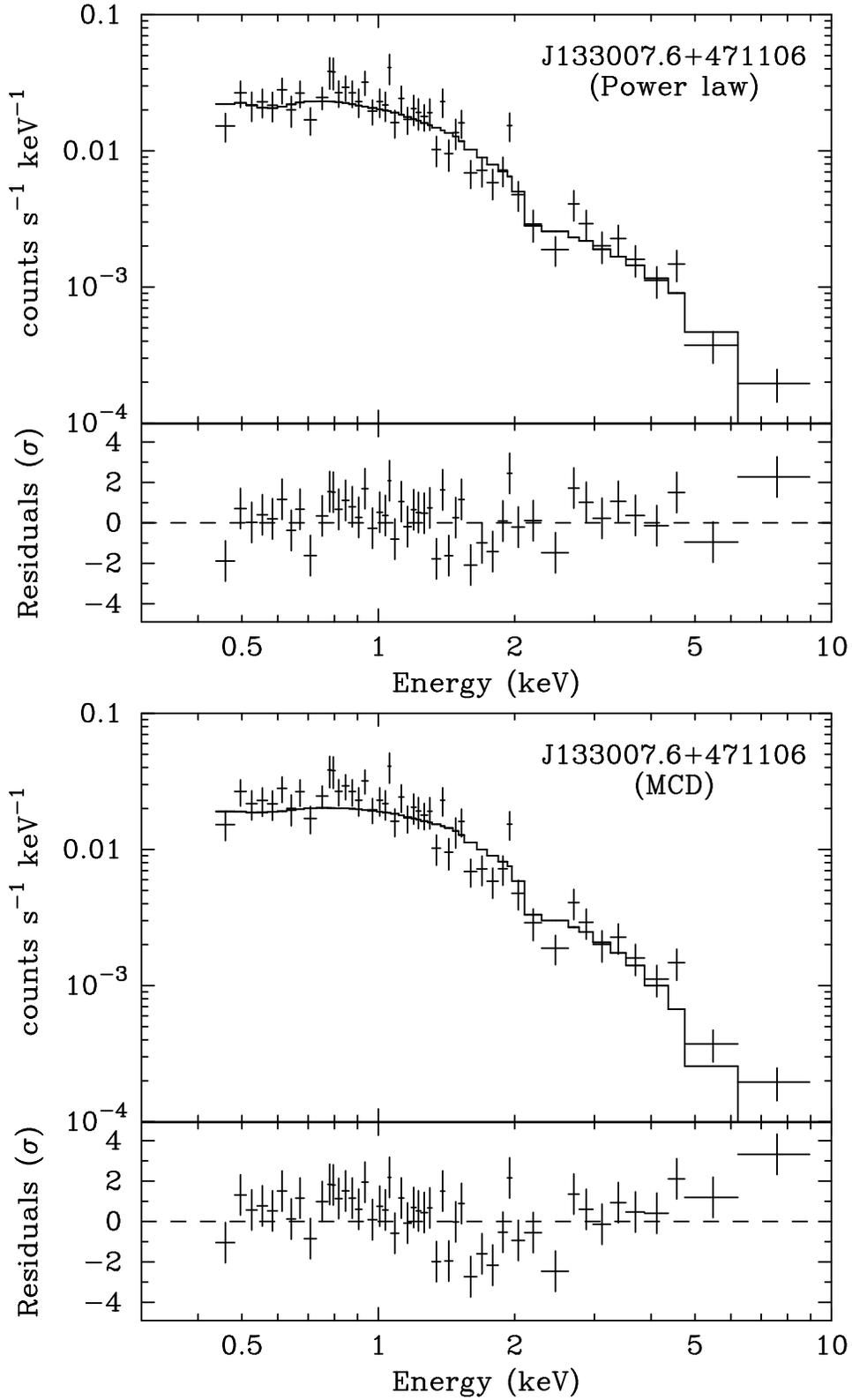

\begin{center}
\includegraphics[scale=0.65,angle=-90]{f8a.ps}

\includegraphics[scale=0.65,angle=-90]{f8b.ps}
\figcaption[]{
X-ray spectrum of NGC 5194 \#82 = CXOM51 J133007.6+471106 in 2001 fitted with
(a) a power law and (b) a MCD model.
}
\end{center}
\end{figure*}

\begin{figure*}[ht]
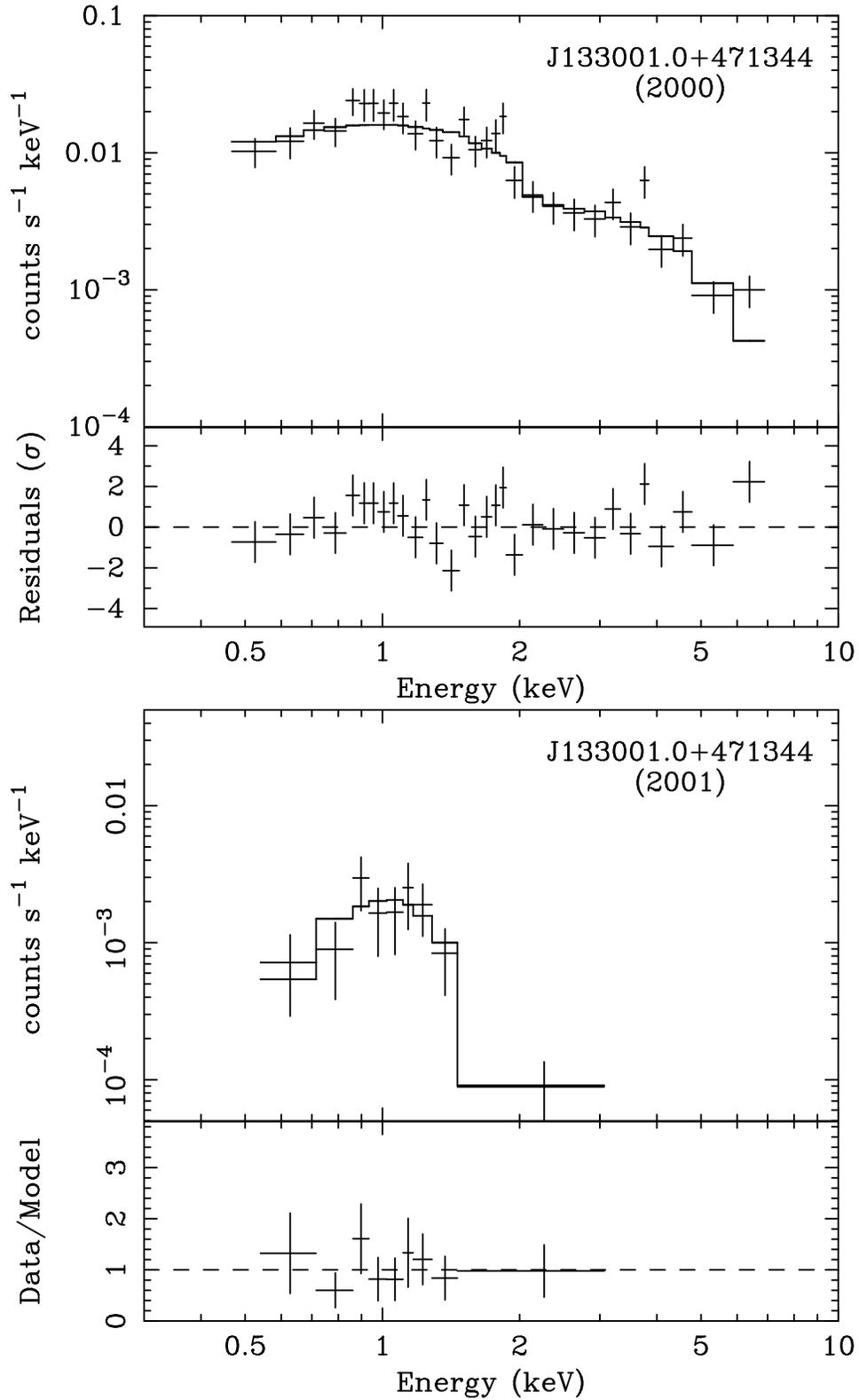

\begin{center}
\includegraphics[scale=0.65,angle=-90]{f9a.ps}

\includegraphics[scale=0.65,angle=-90]{f9b.ps}
\figcaption[]{
X-ray spectra of NGC 5194 \#69 = CXOM51~J133001.0+471344. (a) In 2000
fitted with a power law model. (b) In 2001 fitted with a black body
model.
}
\end{center}
\end{figure*}

\begin{figure*}[ht]
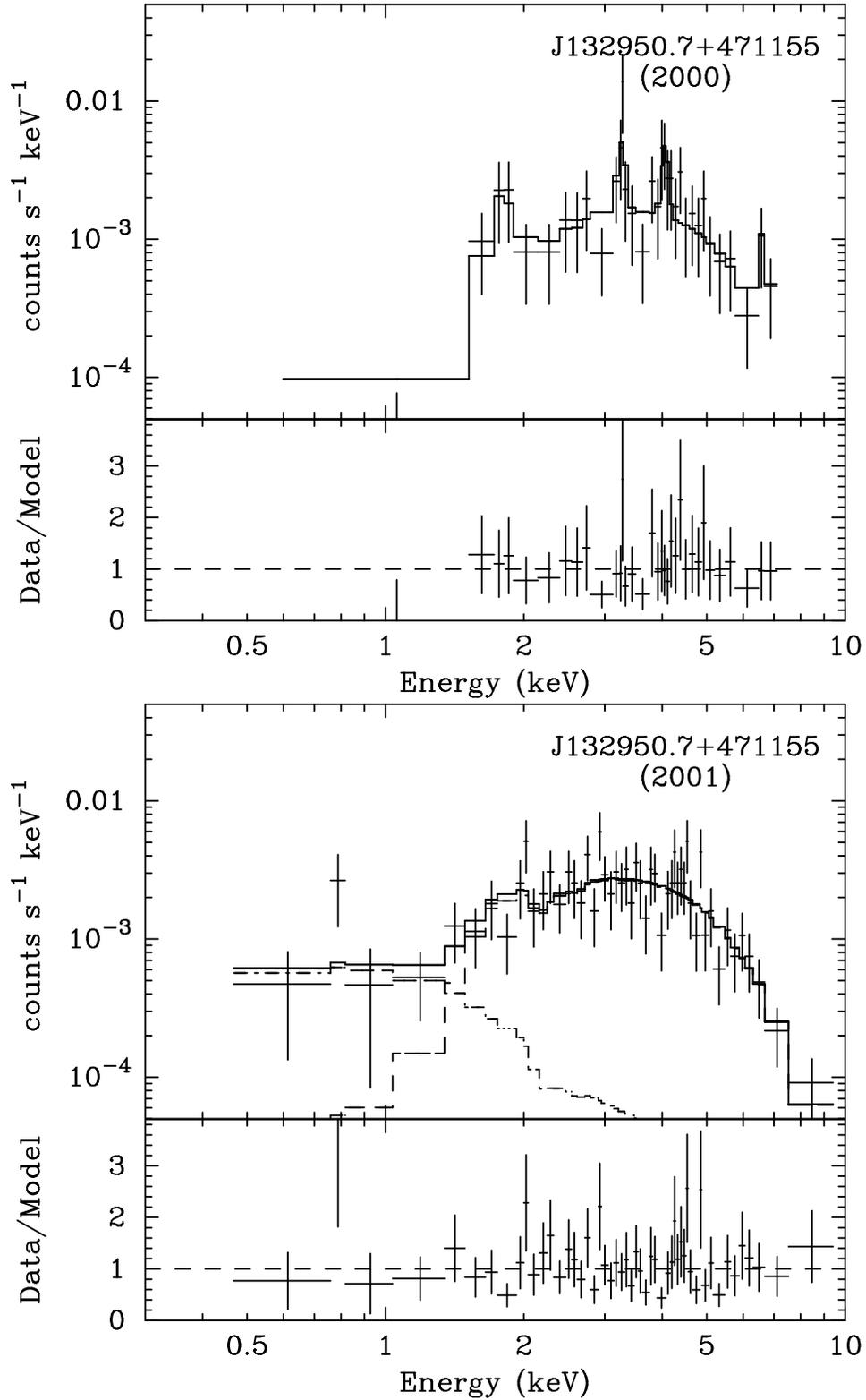

\begin{center}
\includegraphics[scale=0.65,angle=-90]{f10a.ps}

\includegraphics[scale=0.65,angle=-90]{f10b.ps}
\figcaption[]{
X-ray Spectra of NGC 5194 \#26 = CXOM51~J132950.7+471155. (a) In 2000
fitted with a power law and four Gaussians (to represent emission
lines, see Table 8).  (b) In 2001 fitted with a partially covered
power law model.
}
\end{center}
\end{figure*}

\begin{figure*}[ht]
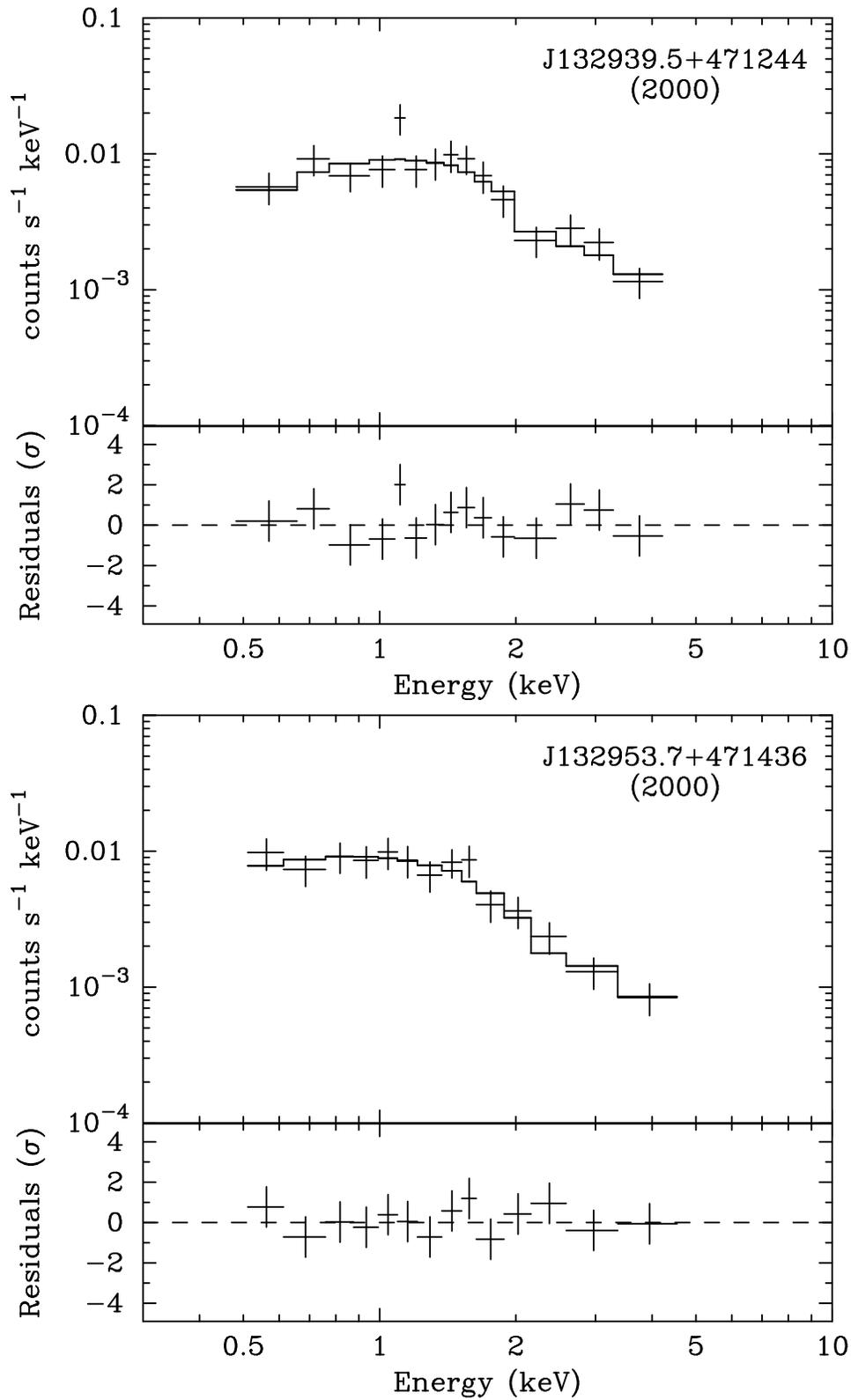

\begin{center}
\includegraphics[scale=0.65,angle=-90]{f11a.ps}

\includegraphics[scale=0.65,angle=-90]{f11b.ps}
\figcaption[]{
X-ray spectra of (a) NGC 5194 \#5 = CXOM51~J132939.5+471244 in 2000
and (b) NGC 5194 \#41 = CXOM51~J132953.7+471436 in 2000, 
each fitted with a MCD model.
}
\end{center}
\end{figure*}

\begin{figure*}[ht]
\begin{center}
\includegraphics[scale=0.65,angle=-90]{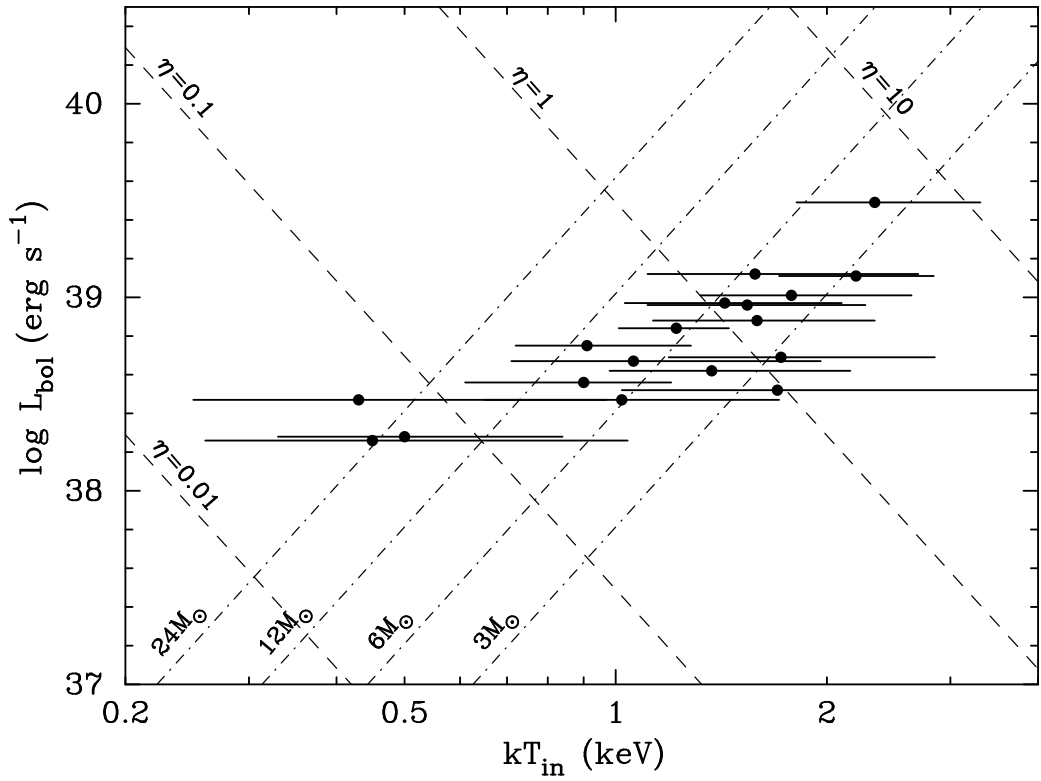}
\figcaption[]{
The relation between the logarithm of the bolometric luminosity, $\log
L_{\rm bol}$, and the inner disk temperature, $kT_{\rm in}$ for a
thin accretion disk. The expected loci of $\log L_{\rm bol}$ and
$kT_{\rm in}$ for constant black hole masses (3, 6, 12, and 24
$M_\odot$) and Eddington ratios ($\eta$ = 0.01, 0.1, 1, and 10) are
shown as the straight dot-dashed and dashed lines,
respectively. $R_{\rm in} = 3 R_{\rm S}$ is assumed.  }
\end{center}
\end{figure*}

\begin{figure*}[ht]
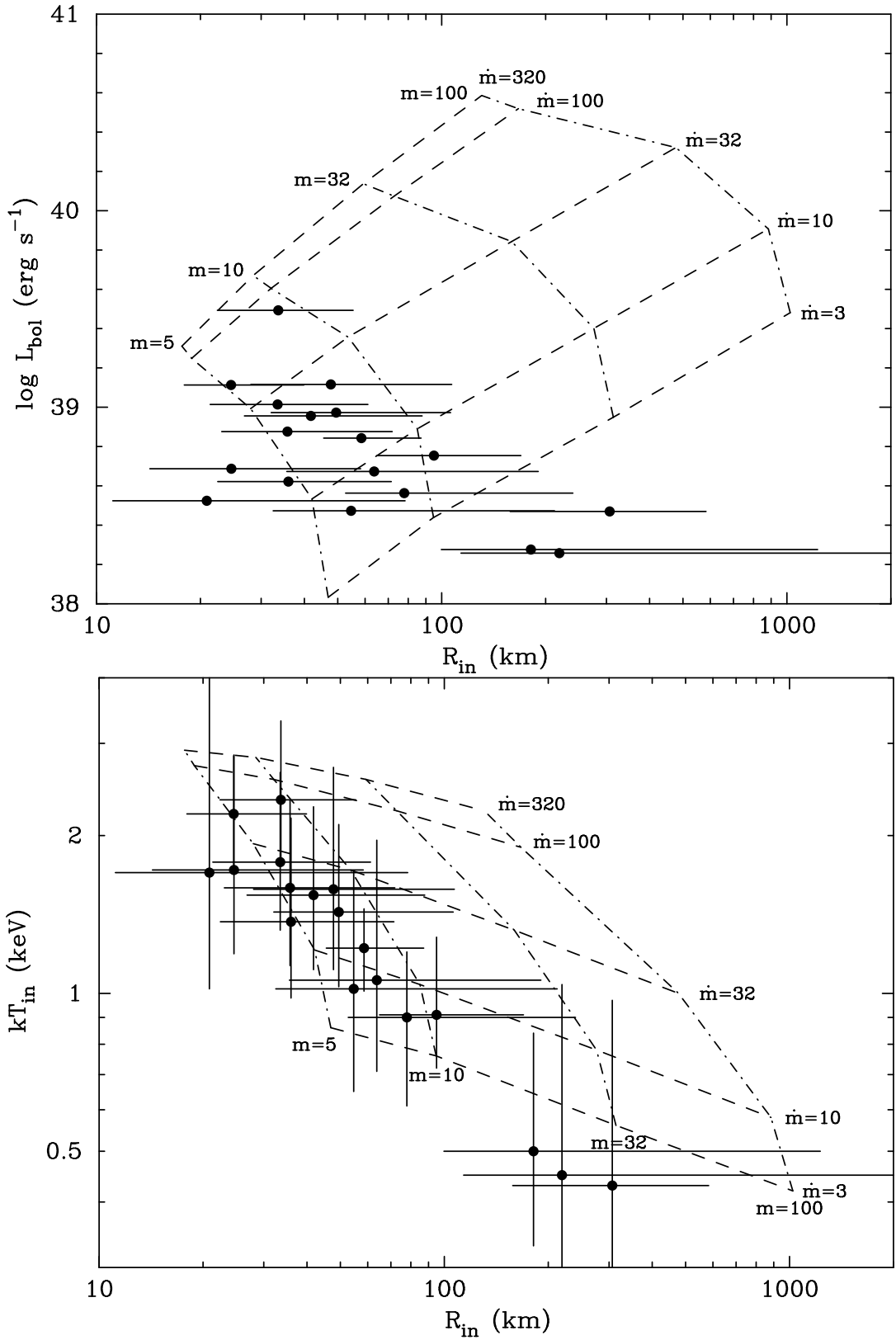

\begin{center}
\includegraphics[scale=0.6,angle=-90]{f13a.ps}

\includegraphics[scale=0.6,angle=-90]{f13b.ps}
\figcaption[]{ 
Comparison between the MCD parameters inferred from the observations
and slim disk model predictions by Watarai et al. (2001).  Dash-dotted
and dashed lines represent constant $m=M/M_{\odot}$ and
$\dot{m}=\dot{M}/(L_{\rm Edd}/c^2)$, respectively.  ({\it a}) $\log
L_{\rm bol}$-$R_{\rm in}$ diagram.  ({\it b}) $T_{\rm in}$-$R_{\rm
in}$ diagram.}
\end{center}
\end{figure*}

\begin{figure*}[ht]
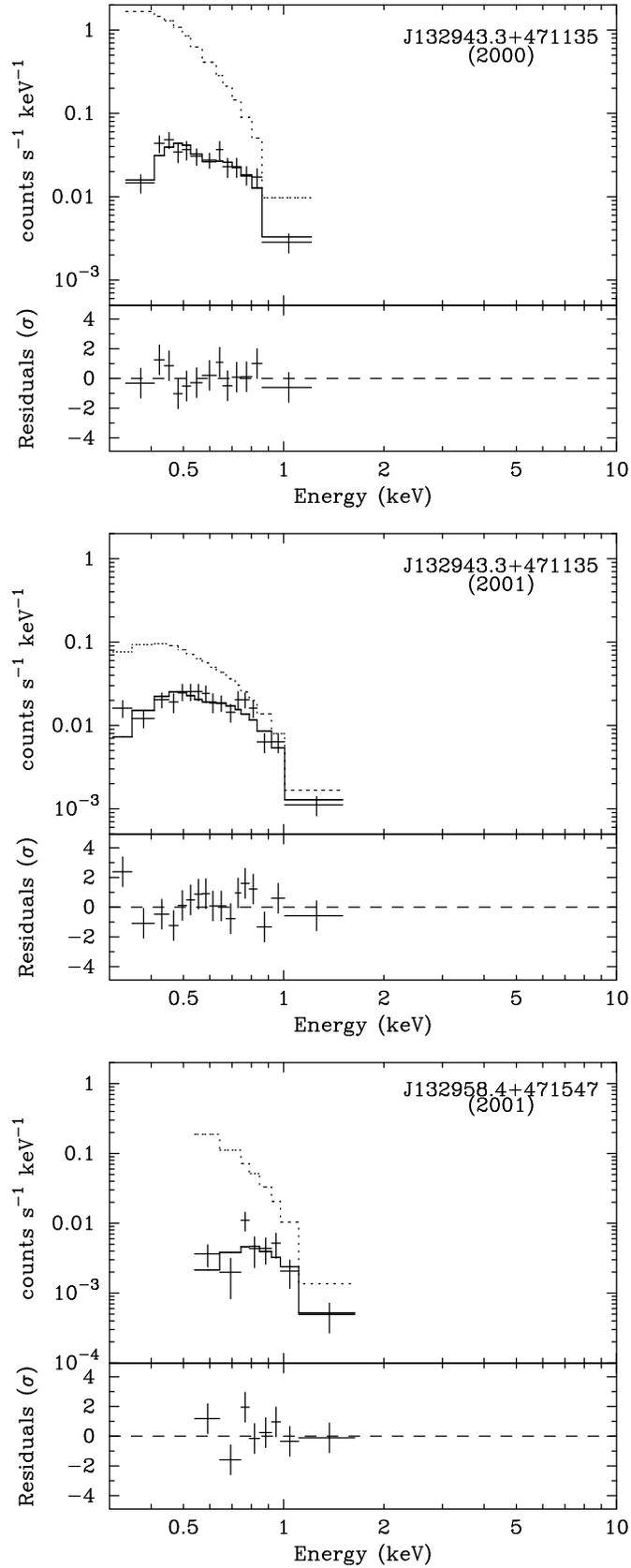

\begin{center}
\includegraphics[scale=0.45,angle=-90]{f14a.ps}

\vspace{3mm}

\includegraphics[scale=0.45,angle=-90]{f14b.ps}

\vspace{3mm}

\includegraphics[scale=0.45,angle=-90]{f14c.ps}
\figcaption[]{
Spectra of super soft sources fitted with
a black body model. Solid and dotted histograms represent
the best-fit model and the model with {\NH} set to be zero, respectively. 
(a) NGC 5194 \#9 in the first observation. 
(b) NGC 5194 \#9 in the second observation. 
(c) NGC 5195 \#5 in the second observation.
}
\end{center}
\end{figure*}

\end{document}